\documentclass[a4paper,11pt,times,astrosym]{aastex631}   

\usepackage[english]{babel}
\usepackage{amsmath}
\usepackage{graphicx}
\usepackage{natbib}

\submitjournal{ApJS}

\shorttitle{Science opportunities for IMAP-Lo observations of ISN He, Ne and O} 

\newcommand{\kms}{~km~s$^{-1}$}

\newcommand{\cmsq}{~cm$^{-2}$}
\newcommand{\persec}{~s$^{-1}$}

\newcommand{\hepri}{He$_\text{pri}$}
\newcommand{\hesec}{He$_\text{sec}$}
\newcommand{\hesum}{He$_\text{sum}$}
\newcommand{\oxpri}{O$_\text{pri}$}
\newcommand{\oxsec}{O$_\text{sec}$}

\newcommand{\hsum}{H$_\text{sum}$}
\newcommand{\nepri}{Ne$_\text{pri}$}

\begin{document}

\title{Science Opportunities for IMAP-Lo Observations of Interstellar Neutral Helium, Neon and Oxygen During a Maximum of Solar Activity}

\correspondingauthor{M.A. Kubiak}
\email{mkubiak@cbk.waw.pl} 

\author[0000-0002-5204-9645]{M.A. Kubiak}
\affil{Space Research Centre PAS (CBK PAN), Bartycka 18a, 00-716 Warsaw, Poland}

\author[0000-0003-3957-2359]{M. Bzowski}
\affil{Space Research Centre PAS (CBK PAN), Bartycka 18a, 00-716 Warsaw, Poland}

\author[0000-0002-9033-0809]{P. Swaczyna}
\affil{Space Research Centre PAS (CBK PAN), Bartycka 18a, 00-716 Warsaw, Poland}

\author[0000-0002-2745-6978]{E. M{\"o}bius}
\affil{University of New Hampshire, Durham, NH, USA}

\author[0000-0002-3737-9283]{N.A. Schwadron}
\affil{University of New Hampshire, Durham, NH, USA}
\affil{Department of Astrophysical Sciences, Princeton University, Princeton, NJ, USA}

\author[0000-0001-6160-1158]{D.J. McComas}
\affil{Department of Astrophysical Sciences, Princeton University, Princeton, NJ, USA}

\begin{abstract}
Direct-sampling observations of interstellar neutral (ISN) species and their secondary populations give information about the physical state of the local interstellar medium and processes occuring in the outer heliosheath.
Such observations are performed from Earth's orbit by the IBEX-Lo experiment on board the Interstellar Boundary Explorer (IBEX) mission.
IBEX ISN viewing is restricted to directions close to perpendicular to the Earth-Sun line, which limits the observations of interstellar species to several months during the year. 
A greatly improved data set will be possible for the upcoming Interstellar Mapping and Acceleration Probe (IMAP) mission due to a novel concept of putting the IMAP ISN detector, IMAP-Lo, on a pivot platform that varies the angle of observation relative to the Sun-Earth line
and the detector boresight.
Here we suggest a 2 yr scenario for varying the viewing angle in such a way that all the necessary atom components can be observed sufficiently well to achieve the science goals of the nominal IMAP mission.
This scenario facilitates, among others, removal of the correlation of the inflow parameters of interstellar gas, unambiguous analysis of the primary and secondary populations of interstellar helium (He), neon (Ne) and oxygen (O), and determination of the ionization rates of He and Ne free of possible calibration bias.
The scheme is operationally simple, provides a good counting statistics, and synergizes observations of interstellar species and heliospheric energetic neutral atoms. 
\end{abstract}

\section{Introduction}
\label{sec:intro}
\noindent
Direct-sampling of interstellar neutral (ISN) species brings crucial information on the physical state and processes operating within the local interstellar matter and in the boundary region of the heliosphere. 
Such observations were pioneered by the Ulysses/GAS experiment \citep{witte_etal:92a}, which observed ISN He and brought the most accurate determination of the interstellar flow velocity available at that time \citep{witte:04}. 
IBEX-Lo \citep{fuselier_etal:09b}, launched in 2008 as one of the instruments on board the Interstellar Boundary Explorer (IBEX) mission \citep{mccomas_etal:09a}, has the ability to observe ISN He, H, D, O, and Ne \citep{mobius_etal:09a, mobius_etal:09b, bochsler_etal:12a,rodriguez_etal:13a,saul_etal:12a}. 
Observations from IBEX-Lo resulted in more accurate estimates of the flow velocity and temperature of ISN He \citep{bzowski_etal:15a, schwadron_etal:15a, mobius_etal:15b, mccomas_etal:15b, swaczyna_etal:22b}; validation that different ISN heavy species flow with the same velocity relative to the Sun based on analysis of ISN O compared with that of He \citep{schwadron_etal:16a}; determination of the Ne/O abundance in the local interstellar matter \citep{bochsler_etal:12a, park_etal:14a}, and the discovery of the secondary populations\footnote{Secondary populations of ISN species are created in the region ahead of the heliosphere, where the plasma and neutral flows decouple because the plasma flow is perturbed by the obstacle created by the heliopause. This decoupling results in charge-exchange collisions between ions and atoms. The products of these collisions are neutralized ions from the perturbed plasma, with kinematic parameters different to those of the unperturbed ISN atoms. Some of these atoms are able to penetrate to 1 au, unlike the other products of the charge exchange collisions, which are ionized ISN atoms. These latter ones form a new plasma population but are unable to penetrate the heliopause.} of ISN He \citep{kubiak_etal:14a, kubiak_etal:16a} and O \citep{park_etal:16a, park_etal:19a}. 
It was also found that the inflow directions of the primary and secondary populations of ISN He mark the plane defined by the velocity vector of the Sun through the interstellar matter and the interstellar magnetic field vector \citep{zirnstein_etal:16b, kubiak_etal:16a, schwadron_etal:16a}. 

More in-depth analysis of observations of ISN He provided an exciting insight into the  interaction of the heliosphere with the ambient interstellar matter and potentially into the physics of the local interstellar medium.
\citet{bzowski_etal:19a} determined the electron density and related parameters of the ionized component of the local interstellar matter.
\citet{swaczyna_etal:23a} suggested that the neutral component of ISN gas is slowed down and heated by elastic collisions in the outer heliosheath, so estimates of the temperature of the local interstellar gas must be revised.
\citet{wood_etal:19a} showed evidence that the temperature of ISN He is anisotropic, and \citet{swaczyna_etal:22c} suggested that the Sun is likely penetrating a region in the interstellar space where the neutral component from two nearby interstellar clouds is mixing, i.e., it is likely in a non-equilibrium state. 

This discussion illustrates the plethora of information that can be retrieved from measurements of the ISN gas. 
However, retrieval of this information has been hampered by limitations in the geometry of available observations. 
While IBEX has provided data from more than one cycle of the solar activity, its viewing geometry is limited to Sun-centered great circles in the sky with the boresight of the sensor almost perpendicular to the radial direction.
This severely restricts the time during each calendar year when the ISN gas is within the instrument field of view and thus can be observed.

\citet{mccomas_etal:18a} detailed the IMAP mission concept that utilizes a sensor IMAP-Lo  mounted on a pivot platform that enables adjusting the angle between the rotation axis and the boresight of the detector, referred to as the elongation angle $\varepsilon$. 
With this, it is possible to adjust the radius of the Sun-centered scanning circle and with that, to perform observations of the ISN species and populations during entire calendar year. 
The ability to vary $\varepsilon$ between 60\degr{} and 160\degr{} extends the time during the year when ISN gas observations can be performed from approximately three to full twelve months.
Effectively it means being able to obtain many independent viewings of the ISN flow beams from different parts of Earth's orbit, instead of just a single ISN map per year from one part of Earth's orbit like IBEX does, which introduces problematic biases and systematic errors. This provides many crucial advantages.
\citet{sokol_etal:19c} suggested that these observation capabilities offer exciting science opportunities: 
\begin{enumerate}
\item improvement of the precision of the flow direction of interstellar neutral species and identifying hypothetical differences between them,
\item unambiguous resolution of the primary and secondary populations of ISN species,
\item investigation of the geometry of the flow of interstellar matter in the outer heliosheath (by analysis of the secondary populations of ISN species),
\item understanding the filtration of the ISN species in the outer heliosheath and inside the heliosphere, and as a result, their ionization state in the ambient interstellar matter,
\item verification of hypothetical departures of the ISN gas from the equilibrium state, 
\item independent determination of the ionization rates of neutral species inside the heliosphere without calibration bias,
\item determining the abundances of Ne, O, and D in the local interstellar medium relative to He and H.
\end{enumerate}

However, the detection mechanism relying on sputtering of negative ions from a special conversion surface by the impacting neutral atoms has important consequences for planning the observations. 
On hitting the conversion surface, some of the ISN H, D and O atoms capture an electron, becoming H$^-$, D$^-$ and O$^-$ ions, which are detected and identified by the instrument. 
Noble gases, like He and Ne, have very low probabilities to form negative ions, and thus are rarely detected directly.
However, when impacting the conversion surface, they more likely sputter negative ions of H, C, and O, which are the main signal source for these species.
As a consequence, it is not possible to easily resolve He and Ne from H and O. 

The primary and secondary populations have inflow velocities that differ by $\sim 10$ \kms, but due to solar gravitational acceleration their impact energies at the detector are very similar (see the Appendix). 
Because the capabilities of the instrument to resolve the energies of the ISN species are limited, especially for species observed through sputtered ions \citep{swaczyna_etal:23b}, it is not possible to resolve the primary and secondary components of a given species by their energies.

Clearly, successfully addressing the science opportunities requires careful studies to identify the times and viewing geometries best suited for gathering the unambiguous data needed for individual science objectives.  
Some of these studies have already been performed.
\citet{schwadron_etal:22a} and \citet{bzowski_etal:22a} explained the correlation of the flow parameters and temperature of the ISN species found in IBEX observations by \citet{bzowski_etal:12a}, \citet{mobius_etal:12a}, and \citet{swaczyna_etal:15a} and suggested prerequisites for the observation geometry of IMAP-Lo that facilitate their removal. 
\citet{bzowski_etal:23a} proposed how to observe ISN He in order to get insight into the ionization rate of this species in the inner heliosphere by coordinated observations of the direct and indirect\footnote{For definitions of the direct and indirect beams, see Section 3.3.1 and Figure 2 in \citet{bzowski_etal:23a}} ISN beams.

This paper belongs to this line of studies. 
We address observations of heavy species, i.e., those for which radiation pressure is negligible: He, Ne, and O \citep{IKL:23a}.
The discussion of H and D will be provided in a future paper (Kubiak et al., in preparation). 

We start with a brief discussion of the simulations used to develop the proposed observation scenarios (Section \ref{sec:simul}).
Subsequently, after presenting a general view of the flow of ISN species from a spacecraft observing from the solar-terrestrial L1 Lagrange point, like IMAP (Section \ref{sec:global}), we study a possible two-year scenario of elongation angle adjustments to address the science opportunities (Section \ref{sec:obsPlan}). 
We focus on the phase of high solar activity, which is expected at the time of the planned IMAP launch in 2025, but many of its considerations are applicable to other time intervals as well. 
However, some adjustments to the proposed scenario might be needed should it be implemented during low solar activity. 

Within the suggested scenario, we focus on several specific topics: observations of He without contribution to the signal from H (Section \ref{sec:HeNoH}), of the primary population of He without the secondary (Section \ref{sec:onlyHepr}), and of the secondary population of He without the primary (Section \ref{sec:onlyHeSc}).
Next, we address ISN Ne and O (Section \ref{sec:NeOx}), in particular how to separate Ne from O given the fact they are difficult to resolve at the detector level (Section \ref{sec:mainNeOx}), how to take advantage of observing Ne in the anti-ram viewing geometry (Section \ref{sec:NeAntiram}) and how to look for the secondary O population (Section \ref{sec:OxSec}). 
We also discuss observations of the indirect Ne beam that may facilitate determination of the ionization rate of Ne (Section \ref{sec:Neindir}).
We close the paper with a summary and conclusions (Section \ref{sec:conclu}).

\section{Simulations}
\label{sec:simul}
\noindent
In this paper, we used the same simulation pool that was used by \citet{sokol_etal:19c} (see their Section 4), supplemented with simulations of the hypothetical secondary population of ISN O.
The simulations comprise the fluxes of various ISN species and their populations as a function of spin angle of the modeled IMAP spacecraft, calculated for Sun-centered circles in the sky with various radii, and integrated over the IMAP-Lo collimator. 
The simulations were performed using the WTPM simulation model \citep{sokol_etal:15b}, with the ionization rates modeled following the observation-based, three-dimensional and time dependent model by \citet{sokol_etal:19a}.
We use the subset of simulations performed for the ionization conditions of 2015, i.e., the high solar activity epoch, because the IMAP launch is scheduled for 2025, i.e., in a similar phase of solar activity. 

The inflow parameters used in the simulations are listed in Section 4 in \citet{sokol_etal:19c}. The ionization rates used, and their evolution in time, are presented in Figure 8 in \citet{sokol_etal:19a}. The contributions to the total rates from individual ionization reactions are presented in Figure 3 in their paper.

We stress that the ionization rate model by \citet{sokol_etal:19a} is not the most up to date available. Since its publication, more actual models have been developed \citep{sokol_etal:20a, porowski_etal:23a}. 
We decided to continue to use the simulations based on the model by \citet{sokol_etal:19a} to maintain homogeneity with the previous papers in this series, mentioned in the Introduction section. 
However, the simulation results we present here cannot be regarded as predictions. On the one hand, this is because the ionization rate model used is not the actual one, and on the other hand, we cannot precisely predict details such as the magnitudes of the ionization rates at the times of the future observations, details of the viewing geometry such as the exact pointing of the spacecraft spin axis, etc.

The simulations are available for the primary and secondary populations of ISN H, D, He, and O, and for the primary population of ISN Ne. 
For these populations, we simulated their fluxes at the virtual IMAP-Lo instrument in the spacecraft-inertial reference frame for every 4-th day of year (DOY) during a calendar year, for the elongation angles $\varepsilon$ of the instrument boresight from the Sun-centered rotation axis of the virtual spacecraft ranging from $\varepsilon = 60\degr${} to $\varepsilon = 180\degr${} with a step of 4\degr, and additionally for $\varepsilon = 90\degr$. 
The flux is calculated averaged over 6\degr{} bins in the spin angle for a given elongation.
For the rationale and presentation of the adopted densities, temperatures, and inflow velocities of individual populations, see Section 4 in \citet{sokol_etal:19c}. 
The parameters adopted for the secondary O population are provided in Section \ref{sec:OxSec}, where opportunities for detection of this population are discussed.

Similarly as in \citet{sokol_etal:19c}, we adopted thresholds for the flux magnitudes and atom energies. These thresholds are an educated guess based on the experience obtained from operating the IBEX-Lo experiment. For He (both populations), the minimum flux is adopted as 100 at \cmsq \persec, and minimum energy at 20 eV in the spacecraft-inertial frame, based on the insight from \citet{galli_etal:15a} and \citet{sokol_etal:15a}. 
For Ne and O, the flux threshold is adopted as 10 at \cmsq \persec.  

\section{Global picture}
\label{sec:global}
\noindent
Atoms of each of the populations of the heavy species follow very similar trajectories inside the heliosphere. 
A virtual detector installed on a Sun-orbiting spacecraft at 1 au, able to image the flux from the entire sky, would observe two diffuse beams of the atom fluxes during each DOY,  for the direct and indirect orbits -- see Figure 4 in \citet{sokol_etal:19c}. 

\begin{figure}
\centering
\includegraphics[width=0.30\textwidth]{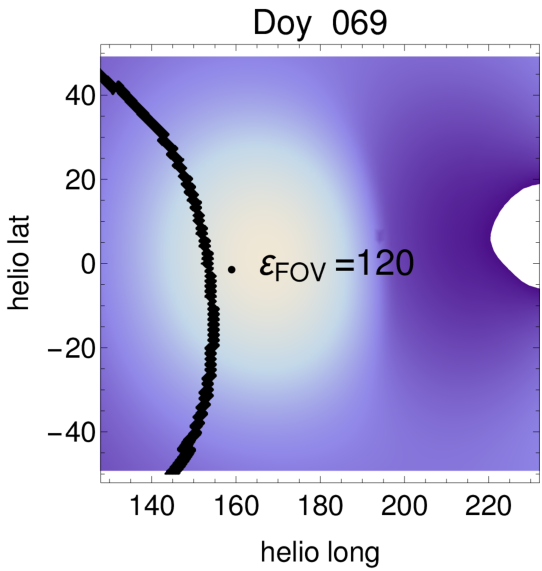}
\includegraphics[width=0.30\textwidth]{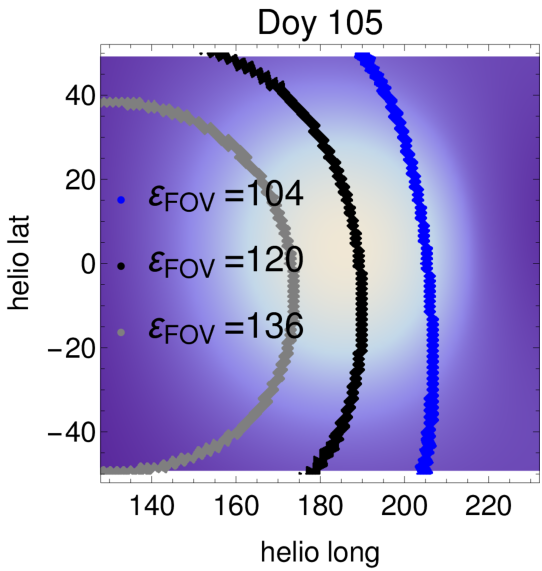}
\includegraphics[width=0.30\textwidth]{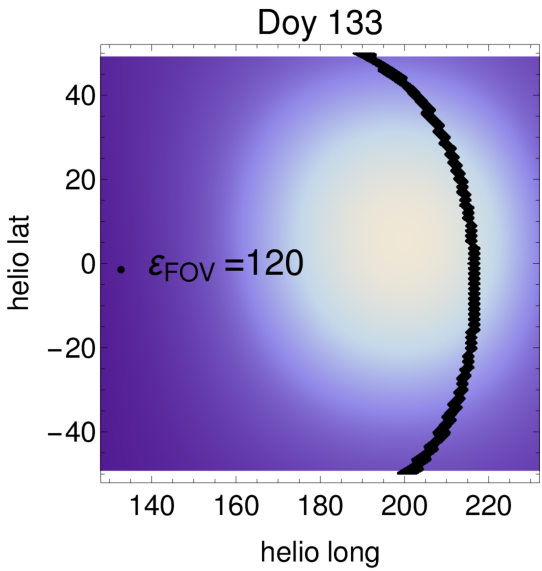}
\caption{
Apparent motion of the beam of ISN He in the spacecraft-inertial frame for $\sim 2$ months during the year, presented in heliographic coordinates. 
The color-coded flux magnitude images can be regarded as daily snapshots of the energy-integrated, spatially resolved differential flux of ISN He for DOYs 69, 105, and 133, from left to right. 
The lines represent portions of Sun-centered scanning circles, with the angular radii (elongation angles $\varepsilon$) listed in the second panel. 
The selected DOYs are identical to those represented as blue dots in the Figure \ref{fig:crossEdoy} and used in the three bottom right-most columns of panels in Figure \ref{fig:crossE120}.
}
\label{fig:mapCross}
\end{figure}

For the primary populations of the heavy species, their respective beam peaks are co-located in the sky, assuming that their inflow velocities are identical.
The difference is in the width of the beams, with those for the heavier species being more narrow than those for the lighter ones, and in the relative heights of the direct and indirect peaks, which depend on the intensity of ionization of a given species in the heliosphere \citep{bzowski_etal:23a}. 
The absolute heights of the peaks depend on the densities of the respective species in the local interstellar matter, and on the ionization losses inside the heliosphere \citep{bzowski_etal:13b}. 
In this work, in line with \citet{sokol_etal:19c}, we used the same densities as those adopted by \citet{bzowski_etal:13b}.

The secondary populations have different inflow velocities and temperatures from those of the corresponding primary species, but they are also expected to have two flux peaks, more diffuse and a little differently located in the sky as compared with their primary-population counterparts. 

The location of the ISN beam in the sky changes with the vantage point, i.e., with the DOY. 
This is illustrated in the sequence of panels presented in Figure \ref{fig:mapCross}, which show the ISN beam as the whitish region that changes its position in the heliographic coordinates in the spacecraft-inertial frame. 
A detector scanning the sky along a Sun-centered circle will cross the ISN beam and register a signal that varies as a function of the spacecraft spin angle. 
For the same vantage location (i.e., for a given DOY), one can use different elongation angles, i.e., the radii of the scanning circle, as illustrated in the second panel of Figure \ref{fig:mapCross} by the colored traces of portions of the scanning circles with various elongation angles.
If the elongation angle is kept unchanged during several days of observations, then the flux vs spin angle will change with time because the locus in the sky of the scanning circle changes daily, and the ISN beam is moving in the sky viewed from the Sun-orbiting spacecraft.
This is exemplified by the black scanning circle in all three panels of Figure \ref{fig:mapCross}.

\begin{figure}
\centering
\includegraphics[width=0.7\textwidth]{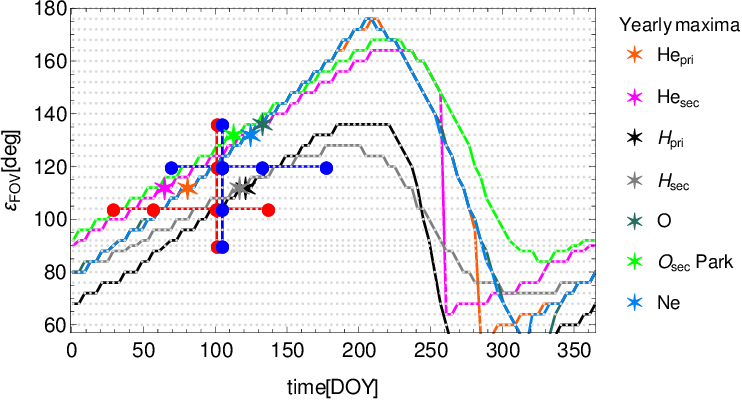}
\caption{
\emph{crossEdoy}
Combinations of elongations and DOYs for which the maxima of the fluxes of ISN species and their populations are expected. 
The gray grid marks the $(\varepsilon, \text{DOY})$ combinations available in the simulation pool. 
The colored lines mark the progression of the positions of ISN beam peaks of the species/populations. 
The large colored asterisks mark the (DOY, $\varepsilon$) combinations for which the fluxes of the respective ISN populations are the largest during the year (see the species legend). 
The red (blue) dots connected with the solid horizontal and vertical lines in respective colors correspond to two alternative sections in the (DOY, $\varepsilon$) space with a fixed either DOY or $\varepsilon$ that offer almost identical fluxes of the ISN species (see text). 
The evolution of the peak maxima for the blue cross for individual atom populations are presented in Figure \ref{fig:crossEdoy2}.
The blue dots represent the days for which we show the fluxes in Figure \ref{fig:crossE120}.
Note that all of the lines shown in the figure appear to be broken because of the overlayed simulation grid. }
\label{fig:crossEdoy}
\end{figure}
\begin{figure}
\centering
\includegraphics[width=0.4\textwidth]{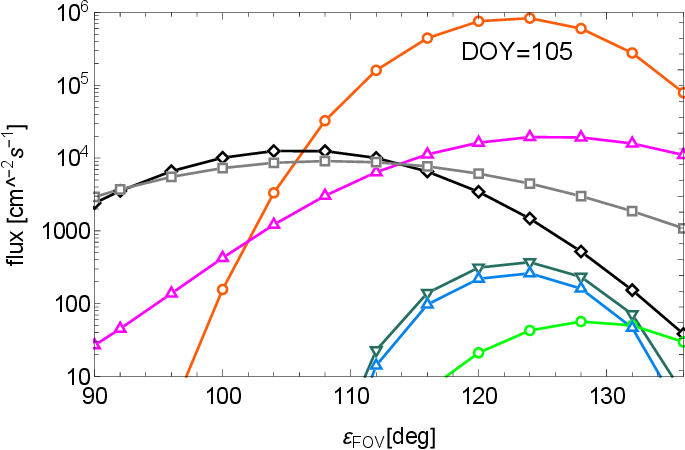}
\includegraphics[width=0.485\textwidth]{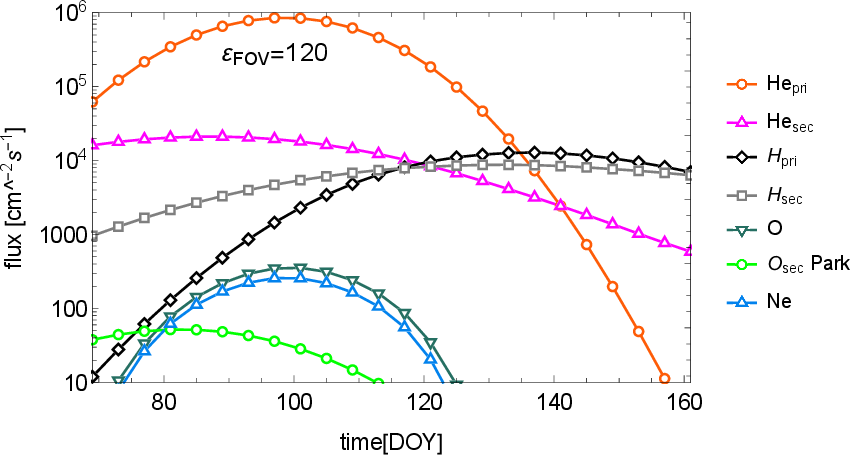}
\caption{
First panel: magnitudes of the spin-angle maxima of the flux as a function of the elongation angle $\varepsilon$ for a fixed DOY 105 and various elongations, i.e., for the vertical bar of the blue cross in Figure \ref{fig:crossEdoy}.
Second panel: magnitudes of the spin-angle maxima of the flux as a function of DOY for a fixed elongation $\varepsilon = 120\degr$, i.e., for the horizontal bar of the blue cross shown in Figure \ref{fig:crossEdoy}. 
} 
\label{fig:crossEdoy2}
\end{figure}

The problem of how to optimally set the instrument to accomplish the science goals -- has five variables: 
\begin{enumerate}
\item the vantage point, parametrized by DOY, 
\item the elongation of the instrument boresight $\varepsilon$,
\item the expected magnitude of the flux of a given species,
\item the impact energies of the atoms, 
\item the spin angle range for which the atoms are observable.
\end{enumerate}

Accomplishing all of the science goals requires finding a compromise between optimum sampling of an individual species population and satisfying prerequisites for the other species populations. 
Optimizing the science returns requires finding an optimum subset of (DOY, $\varepsilon$) pairs while taking into account the other three dimensions in the problem space, i.e., the spin angle, energy and flux magnitude aspects.

To characterize the ISN flux distributions in the sky, it is desirable to obtain cuts through them in two  directions. 
For a detector scanning Sun-centered sky strips, like IMAP-Lo, one of these cuts is naturally obtained due to the spinning of the spacecraft. 
A priori, a cut in another direction can be obtained in two alternative ways: either by adjusting the elongation angle during one observation day, or by maintaining the elongation angle fixed for a certain time interval and waiting for the atom beam to swipe across the scanning circle. 
The first of these two strategies is not feasible for IMAP-Lo because of operational reasons. We have performed simulations which suggest that using the second of these strategies provides almost perfectly equivalent results. 
\begin{figure}
\centering
\includegraphics[width=0.24\textwidth]{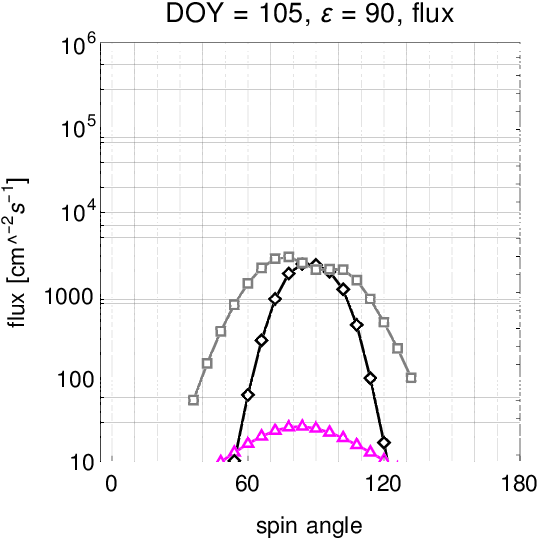}
\includegraphics[width=0.24\textwidth]{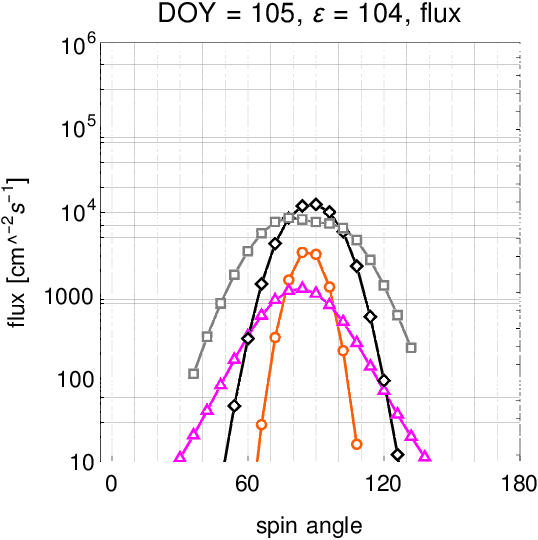}
\includegraphics[width=0.24\textwidth]{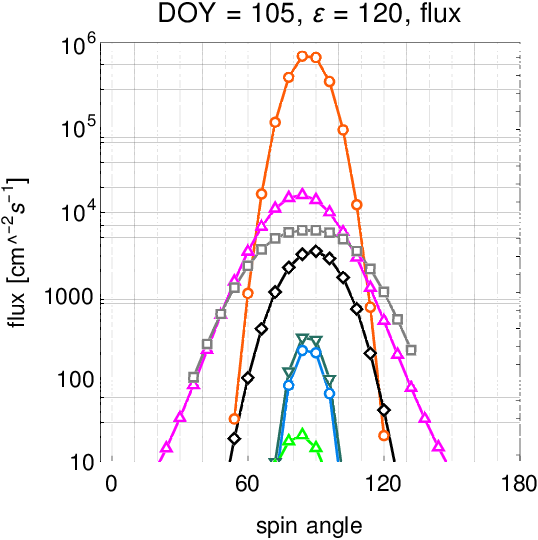}
\includegraphics[width=0.24\textwidth]{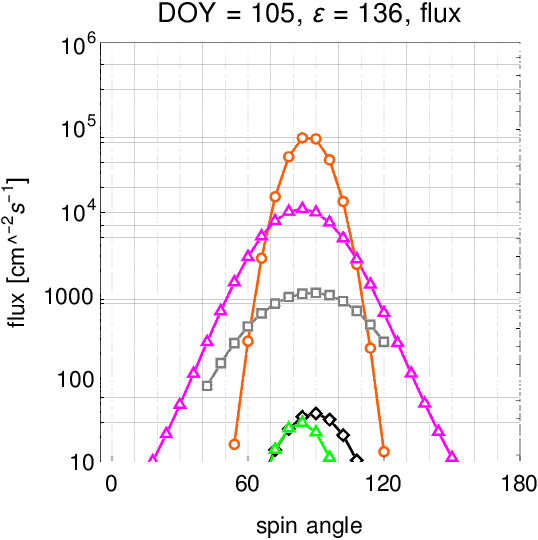}

\includegraphics[width=0.24\textwidth]{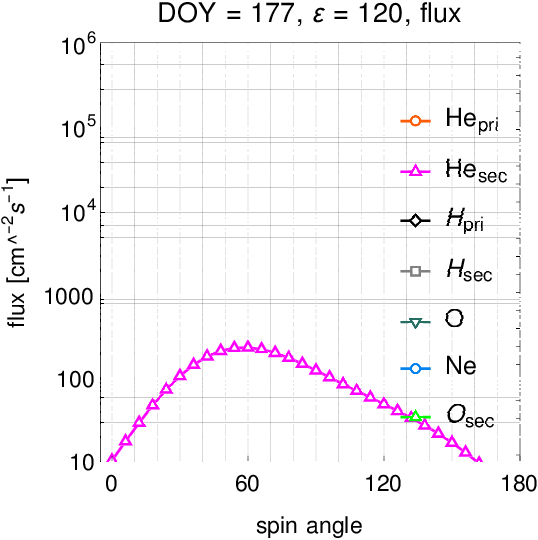}
\includegraphics[width=0.24\textwidth]{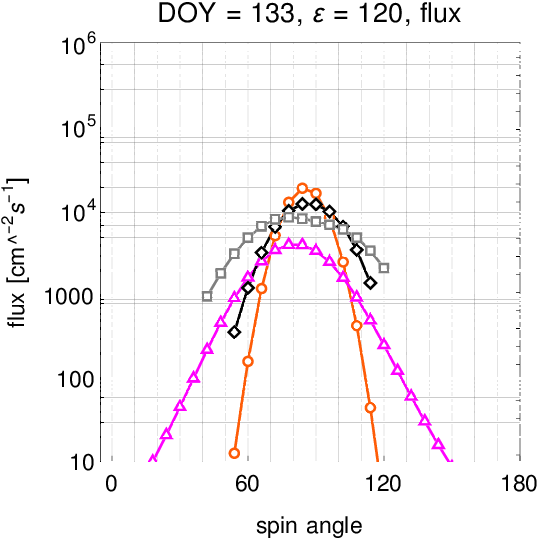}
\includegraphics[width=0.24\textwidth]{figures/doy105.elong.120.He.cross.eps}
\includegraphics[width=0.24\textwidth]{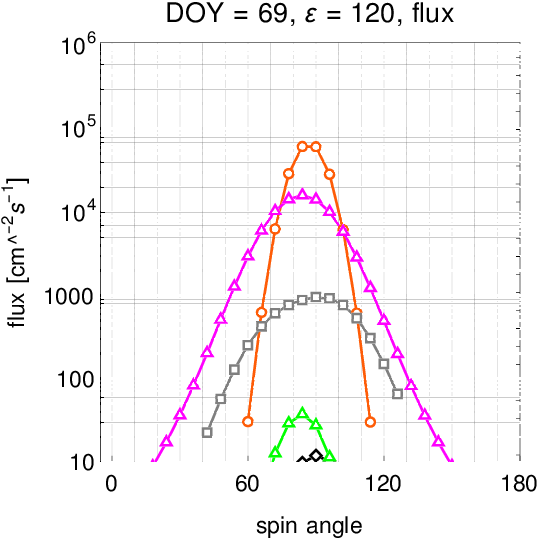}
\caption{
Expected fluxes of ISN species  for the two perpendicular cuts through the (DOY, $\varepsilon$) space, displayed Figure \ref{fig:crossEdoy} as the blue cross. The upper row presents the fluxes for a number of elongations for a fixed DOY = 105 (the dots in the vertical bar of the cross). 
The lower row shows the fluxes for a number of DOYs for a fixed elongation $\varepsilon = 120\degr$ (the dots in the horizontal bar of the cross). 
The figure demonstrates that even though the two rows correspond to different loci in the (DOY, $\varepsilon$) space, the fluxes are very similar to each other within the panel columns. 
 }
\label{fig:crossE120}
\end{figure}

We compared flux simulations for both of the populations of ISN H, He, O, as well as the primary population of Ne, for two alternative sequences of (DOY, $\varepsilon$) series. 
The first of them corresponds to the $\varepsilon$ scan: DOY 105, $\varepsilon \in \{90\degr, 104\degr, 120\degr, 136\degr \}$, and is represented by the vertical bar of the blue cross shown in Figure \ref{fig:crossEdoy}.
The other scan is obtained by continuing observation with a fixed elongation for a certain range of DOYs: $\varepsilon = 120\degr, \text{DOY} \in \{177, 133, 105, 69 \}$, which  represents the horizontal bar of the blue cross in Figure \ref{fig:crossEdoy}. The absolute yearly maxima of the ISN fluxes are marked with the colored asterisks.
The yearly sequence of the elongation angles needed to view the maxima of the beams of individual species are presented as colored lines. 

The spin-angle maxima of the fluxes simulated for the horizontal and vertical bars of the blue cross in Figure \ref{fig:crossEdoy} are presented in the first and second panel of Figure \ref{fig:crossEdoy2}, respectively. 
The shapes of the lines in the (DOY, flux) space in the left panel are very similar to the equivalent lines in the (elongation, flux) space in the right panel of the figure.
This shows that information obtained from these two alternative sections of the ISN beams in the sky is equivalent to each other.

Detailed illustration of the fluxes vs spin angles for these scans are shown in Figure \ref{fig:crossE120}. The first row of panels corresponds to the varying $\varepsilon$ strategy, the second row to the fixed $\varepsilon$ approach. 
Note that the third panels from the left in both rows are identical and correspond to the crossing point of the blue lines shown in Figure \ref{fig:crossEdoy}.
Clearly, the spin-angle runs of the fluxes shown in the two rows of panels in Figure \ref{fig:crossE120} are very similar to each other. 
We verified that very similar results and conclusions are obtained for the beam cuts illustrated by the red cross in Figure \ref{fig:crossEdoy}.

Based on this insight we conclude that scanning across the ISN beams by maintaining the elongation angle fixed for a number of weeks does not result in an information loss in comparison with a strategy where the local ISN flux maximum is tracked by a daily adjustment of the elongation angle.
The step-like elongation adjustment strategy with relatively small number of the yearly steps permits the observation of all species and their populations, which is impossible to accomplish with the strategy of following a beam maximum with daily adjustments of the elongation angle. 

In the following section, we will demonstrate how to achieve all of the goals discussed in the paper within a homogenous and operationally simple system of adjustments of the instrument elongation during two calendar years of operations. Subsequently, we will focus on interesting aspects of studies of ISN He (the primary and secondary populations), and the heaviest species, i.e., O and Ne. We will consider (DOY, $\varepsilon$) combinations most suitable to address these topics.  

\section{Best times and elongation angles for observations of ISN gas}
\label{sec:obsPlan}
\noindent
IMAP will be a spinning spacecraft in a orbit around the solar-terrestrial L1 point, with the rotation axis directed 4\degr{} to the right of the Sun.
Re-adjustments of the rotation axis are planned to be performed daily. 
This is different to IBEX, which is located in a highly-inclined Earth orbit, and its spin axis adjustment twice per revolution, approximately every 4.5 days \citet{mccomas_etal:11a}.
Observations of the ISN species by the IMAP-Lo instrument will be combined with observations of heliospheric energetic neutral atoms (ENAs).
The instrument has settings that optimize detection of neutral atoms in several energy bands.
These settings will be sequenced during each day of observations, so that every day all of the energy bands are scanned.
Hence, in the absence of an energy threshold for detection, it would be possible to observe ISN atoms of all species every day. 

\begin{figure}
\centering
\includegraphics[width=0.75\textwidth]{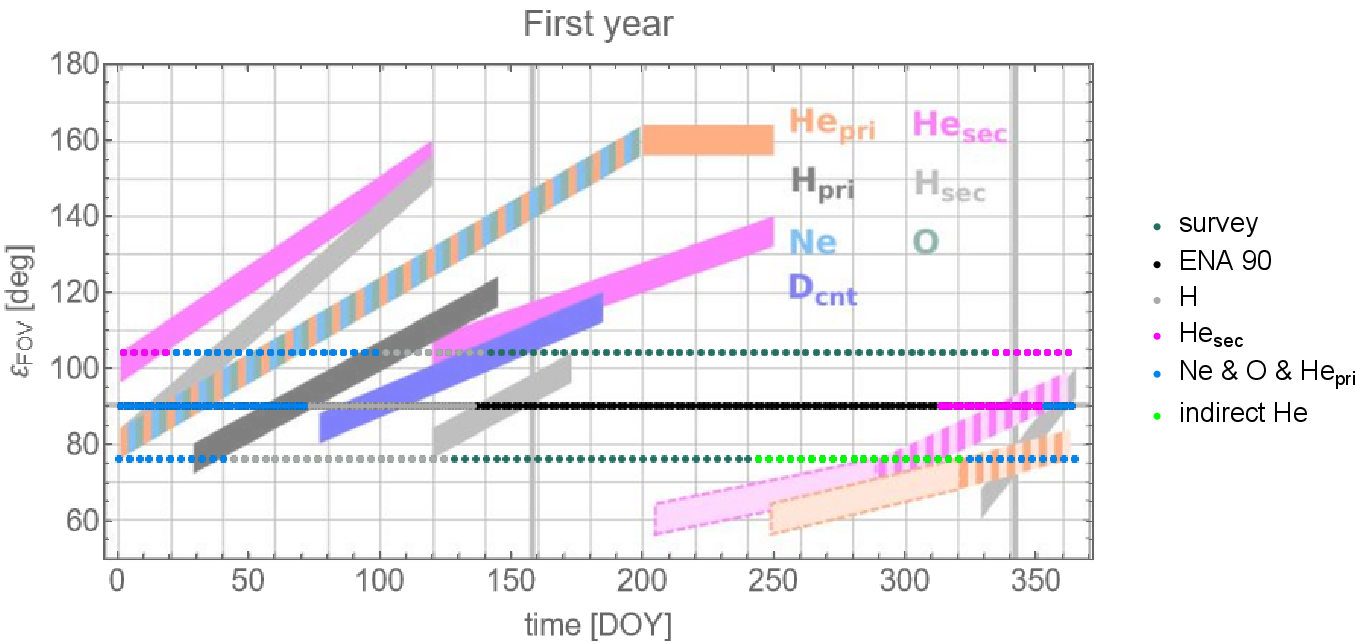}
\includegraphics[width=0.75\textwidth]{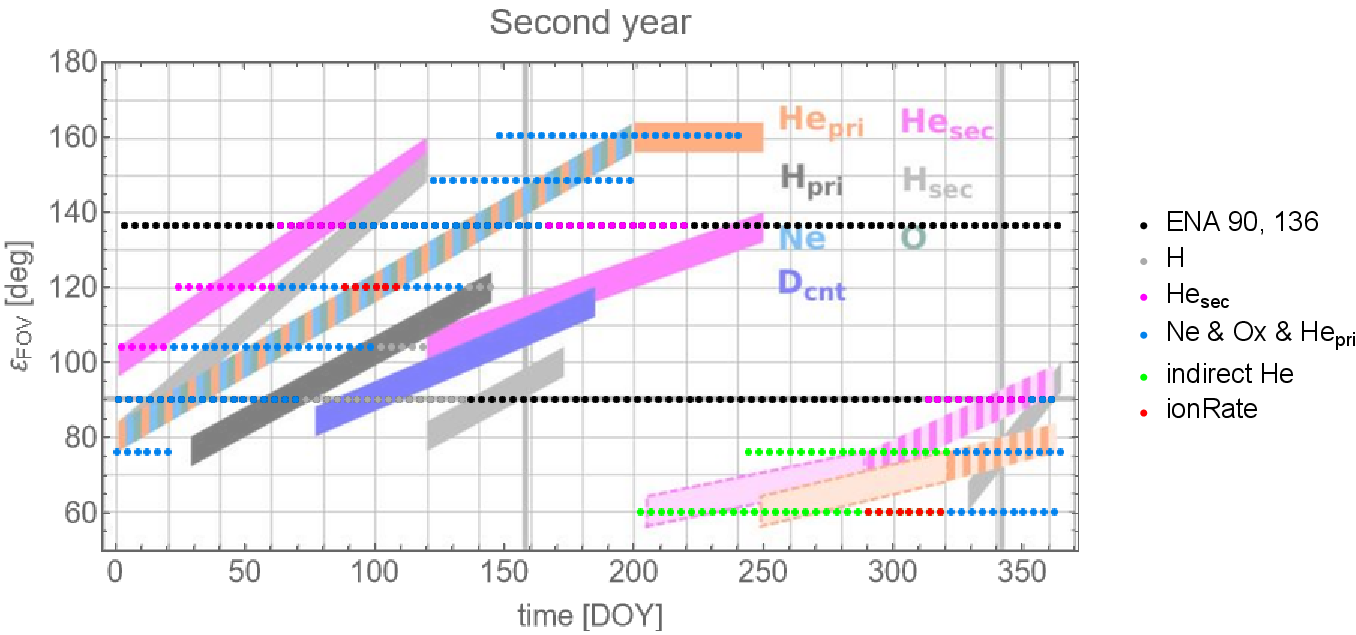}
\caption{
Summary of the elongation settings to maximize science opportunities for the ISN gas studies with IMAP-Lo- to accomplish within two years of operations. 
The upper and lower panel represents the settings for the first and second year, respectively. 
The colors mark the ISN observation campaigns discussed in the text; the corresponding legends are displayed next to the panels.  
To provide a wider context, the background in both of the panels is adapted from the upper panel of Figure 15 in \citet{sokol_etal:19c}.}
\label{fig:obsPlan}
\end{figure}

IMAP-Lo is installed on a pivot platform that enables setting the elongation angle $\varepsilon$ within a range of (60\degr, 160\degr) relative to the spacecraft spin axis.
During the first year of science operations, the range of elongation settings is planned to be restricted to a range of (75\degr, 105\degr).
Because of the available simulation grid, we represent these limiting settings by $\varepsilon = 76\degr$ and $\varepsilon = 104\degr$. 
During the second and subsequent years of operations, these restrictions will be removed.

Accomplishing some of the science goals requires prolonged observations distributed in time during the year. 
For example, \citet{bzowski_etal:22a} showed that accomplishing science goal (1) requires performing observations of the ISN species distributed along a long arc of the Earth orbit around the Sun. 
\citet{sokol_etal:19c} discussed the (DOY, $\varepsilon$) combinations optimized for tracking the flux maxima of individual species and elongations. 
However, it is not possible to perform observations to accomplish all of the science goals following this strategy: optimizing for one species/population compromises some others.
In this paper, we suggest a scenario for adjusting the elongation angle that enables fulfilling all of the science goals without relying on peak tracking.
In other words, \citet{sokol_etal:19c} showed what is there to see, and we show how to see it all within two years. 
Extending the observation period past the two years enables improving the statistics, and monitoring time variations, as it was done on IBEX \citep{swaczyna_etal:22b}.


The suggested scenario of elongation angle settings covers two years.
During this time, the science goals discussed in \citet{sokol_etal:19c} can be successfully accomplished.
The elongation setting scenario is devised based on several prerequisites:
\begin{itemize}
\item the elongation angle is set once for the entire observation day (i.e., for an interval when the pointing of the spacecraft spin axis is fixed); 
\item half of the available observation days are optimized for ISN observations, the other half for the ENAs;
\item energy stepping is identical for the ISN and ENA days and covers the entire available energy range;
\item the elongation settings form carefully placed horizontal lines in the (DOY, $\varepsilon$) space 
\item the elongation settings are varied periodically, with the period of the elongation cycle equal to four DOYs, so that one cycle comprises days A, B, C, and D;
\item within one elongation cycle, there are up to two different settings optimized for the ISN observations and up to two settings for the ENAs;
\item the ISN and ENA days are arranged within the elongation cycle in an alternating pattern, so that days A and C during a cycle are used for ISN observations, and days B and D for the ENAs.
\item the elongation settings for ENA observations enable direct comparison with IBEX on the one hand, and with IMAP-Hi on the other hand, which has two detectors with the elongation angles equal to 90\degr{} and 135\degr.
\end{itemize}

The suggested elongation setting scenario is summarized in Figure \ref{fig:obsPlan}, with the two panels corresponding to the first and second data collection years. 
The sequence can start on any DOY.
In this way, the suggested scenario does not depend on the actual start date of the science operations.
For example, if science operations begin on DOY 186, we suggest to perform the observations according to the schedule presented in the first panel of Figure \ref{fig:obsPlan} starting from DOY 186 until the last day of the calendar year, and on the first day of the subsequent calendar year to continue with the settings listed in the first panel until DOY 185. 
Subsequently, we switch to the elongation setting scenario presented in the second panel of Figure \ref{fig:obsPlan}, starting from DOY 187, etc.

The scenario for the first year minimizes the range of motions of the pivot platform. 
The elongation angle is alternated between 75\degr, 90\degr, and 105\degr.
For the second year, the elongation angle for the ISN days always follows two parallel step-like lines in the (DOY, $\varepsilon$) space. 
This enables performing several passages of the ISN beam through the field of view.
In addition, there are two parallel lines at $\varepsilon = 90\degr$ and 135\degr, optimized for ENA mapping, but also very useful for ISN observations.

In the background of the panels in Figure \ref{fig:obsPlan}, we show the content of the upper panel of Figure 15 in \citet{sokol_etal:19c} to demonstrate that the proposed elongation settings enable accomplishing the science goals discussed in their paper and maximize the science return of the IMAP-Lo instrument. 

For better clarity, we discuss several topical observation campaigns within our suggested scheme. 
Below, we list the campaigns and present the prerequisites for the expected fluxes. 
When defining the times to change the elongation, we often decide to end a given campaign and begin another one even when the prerequisites for the first one are still fulfilled, based on the objective of maximizing the science return in all of the aspects. 
The topical campaigns are the following:
\begin{itemize}
\item\textbf{ Heavy Species Campaign} (blue color in Figure \ref{fig:obsPlan}), optimized for investigating the flow parameters of the primary populations of the heavy species He, Ne, and O (\hepri, \nepri, \oxpri, respectively). 
The campaign begins when the flux of the secondary helium population \hesec{} becomes lower than that of \hepri{} (\hesec{} $<$ \hepri), and ends when \hesum{} = \hepri{} + \hesec $<$ \hsum. 
For high elongations, when there is no H left in the signal because of an insufficient impact energy, the campaign ends when the flux of \nepri{} or \oxpri{} drops below the adopted threshold of 10 atoms \persec \cmsq.
\item \textbf{Secondary Helium Campaign} (magenta color in Figure \ref{fig:obsPlan}), optimized for characterization of the secondary population of ISN He, begins when \hesec $>$ 100 atoms \persec \cmsq{} and ends when \hesec{} $<$ \hepri{} or -- as for $\varepsilon = 136\degr$ after DOY 150 -- the flux fades out. 
\item \textbf{Hydrogen Campaign} (gray color in Figure \ref{fig:obsPlan}), optimized for observations of ISN H, begins when \hsum{} $>$ \hesum{} and ends when the H flux fades out, mostly because of insufficient impact energy. For $\varepsilon > 136\degr$, the energy is anyway too low for detection. 
\item \textbf{Indirect Beam Campaign} (green color in Figure \ref{fig:obsPlan}) involves the smallest available elongations, starting from $\varepsilon = 60\degr$ and is mostly aimed at observations of the indirect beams of \hepri{} and \hesec. Hopefully, also pure Ne can be observed, with expected negligible contribution from \oxpri. In this campaign, we are probing the slopes of the cone of ISN He, gravitationally focused behind the Sun.
\item \textbf{Ionization Rate Campaign} (red color in Figure \ref{fig:obsPlan}), aiming at determination of the ionization rate of ISN He and possibly ISN Ne. Based on \citet{bzowski_etal:23a}, observations of the direct and indirect beams of ISN He are needed for the DOYs when the impact energies of the direct and indirect beams are close to each other.  
\item \textbf{ENA Campaign} (black color in Figure \ref{fig:obsPlan}), with the settings of $\varepsilon = 90\degr${} during the first year, and alternately 90\degr{} and 135\degr{} during the second year, is devised to map the entire sky similarly as it is done on IBEX (for $\varepsilon = 90\degr$) and will be done by IMAP-Hi (for $\varepsilon = 90\degr$ and 135\degr), but many of the observation days will also be useful for ISN studies.
\end{itemize}

In this paper, we dwell mostly on ISN He, Ne, and O.
Discussion of ISN H and D will be presented in a separate paper \citep[in preparation]{kubiak_etal:23b}. Here we mention them only when needed to better understand the rationale behind the elongation setting scenario.

For the first year, assuming that observations start on DOY = 1, the observation cycle is (DOY = A, $\varepsilon = 76\degr$), (DOY = B, $\varepsilon = 90\degr$), (DOY = C, $\varepsilon = 104\degr$), and (DOY = D, $\varepsilon = 90\degr$). This suggested scheme is presented in the first panel of Figure \ref{fig:obsPlan}.

During the second year, we suggest the (DOY, $\varepsilon$) scheme presented in the second panel of Figure \ref{fig:obsPlan}. The ENA days, i.e., days B and D in the 4-day observation cycle, alternate between $\varepsilon$ = 90\degr{} for day B and $\varepsilon = 135\degr${} for day D. The $\varepsilon$ settings for the ISN days (i.e., days A and C in the proposed 4-day cycle) are summarized in Table \ref{tab:Y2}. 

\begin{deluxetable}{|ccr| ll|r|}[!h]
\tablecaption{\label{tab:Y2} \emph{tab:Y2} Days preferable to observe ISN during the second year IMAP-Lo observations.}
\tablehead{ Case no & day in cycle & $\varepsilon$ & start DOY & end DOY & change to $\varepsilon$}
\startdata
1 & A &  76\degr & 1     & 21  & 120\degr \\
2 & C & 104\degr & 3     & 119 & 148\degr \\
3 & A & 120\degr & 25    & 145 & 160\degr \\
4 & C & 148\degr & 123   & 199 &  60\degr \\
5 & A & 160\degr & 149   & 241 &  76\degr \\
6 & C &  60\degr & 203   & 363 & 104\degr \\
7 & A &  76\degr & 245   & 365 &  
\enddata
\end{deluxetable}

For the ISN days, the change from $\varepsilon = 76\degr${} to $\varepsilon = 120\degr${} (Case 1) results in abandoning good H measurements in a region free from ISN He for the sake of the secondary He population in a region free from the primary He. 
Both of these measurements for the DOYs 20--50 would be valuable, but because the elongation $\varepsilon = 76\degr${} is maintained throughout the first year and a good statistics for ISN H is obtained, it is more favorable to use the available time for observations of the pure secondary population of ISN He. 

The change from $\varepsilon = 104\degr$ to $\varepsilon = 148\degr${} (Case 2 in Table \ref{tab:Y2}) is mostly an attempt to look for the secondary O. 
For $\varepsilon = 104\degr$, DOYs 120--130, we still have a good signal from ISN H without ISN He, but these observation will be performed during the first year. 
Therefore, an opportunity to look for the poorly known secondary O, which is expected to be visible during this time for $\varepsilon = 148\degr$, seems more valuable. 

Switching from $\varepsilon = 120\degr$ to 160\degr{} (Case 3 in Table \ref{tab:Y2}) is due to the expected fading of the signal for $\varepsilon = 120\degr$. 
In this region, only ISN H is present, but its energy drops below the detection threshold about DOY 145. 
But during this time, and also earlier, there is a strong signal from ISN He for greater elongations, and shortly after the planned switching to $\varepsilon = 160\degr$, also from ISN Ne, and both populations of ISN O will likely appear. 

The subsequent two elongation changes, from $\varepsilon = 148\degr${} to $\varepsilon = 60\degr${} and from $\varepsilon = 160\degr${} to $\varepsilon = 76\degr$ (cases 4 and 5, respectively) are due to the expected disappearance of the signal from the heavy species at large elongations, and the appearance of the indirect beam at the smaller elongations. 

A presentation of the simulated fluxes for all 4-day cycles for both of the years is offered in the Appendix.


\section{Helium}
\label{sec:He}
\noindent
Helium is the second most abundant species in the local interstellar medium, but because it is very little sensitive to radiation pressure \citep{IKL:23a} and only weakly ionized inside the heliosphere \citep{bzowski_etal:13b}, at 1 au it is the most abundant species. 
ISN He is the main component of the ISN atom signal for IBEX-Lo and IMAP-Lo. 
Measurements of ISN He are an important source of information on the physical state of the matter in the immediate solar Galactic neighborhood. 
Analysis of the relation of the primary and secondary ISN He populations allows for investigation of deformation of the heliosphere from axial symmetry due to the pressure of interstellar magnetic field. 

\citet{sokol_etal:19c} pointed out the regions in the (DOY, $\varepsilon$, flux, energy) space where a measurable ISN He signal is expected. Here, we investigate in greater detail where the science return of ISN He observations will be the largest within a homogeneous observation scenario.

The planned measurements of ISN He by IMAP-Lo have been discussed in terms of breaking the inflow parameter correlation \citep{schwadron_etal:22a, bzowski_etal:22a} and calibration bias-free determination of its ionization rate based on observations of the direct and indirect beams \citep{bzowski_etal:23a}. 
Both of these are included in the observation campaign of the heavy ISN species, marked with the blue and red colors in Figure \ref{fig:obsPlan}. 
The expected flux vs spin angle variations for some selected (DOY, $\varepsilon$) combinations are presented in Figures~\ref{fig:HePrSc}, left panel, and in Figure \ref{fig:crossE120}. 
For the conditions shown in these figures, the instrument field of view contains both populations of ISN He as well as those of ISN H, Ne, and O, largely for the same spin angles. 

To facilitate in-depth analysis, it would be helpful to have observations of each of the two He populations with as little contribution to the signal from the other one as possible, i.e.,  pure primary and pure secondary population signals.
Since it is challenging to unambiguously assign individual events either to He or H, it is also desirable to have measurements of He with the contribution from H minimized.

In the following subsections, we will discuss opportunities to achieve all of these goals.

\subsection{He without H}
\label{sec:HeNoH}

\begin{figure}
\centering
\includegraphics[width=0.24\textwidth]{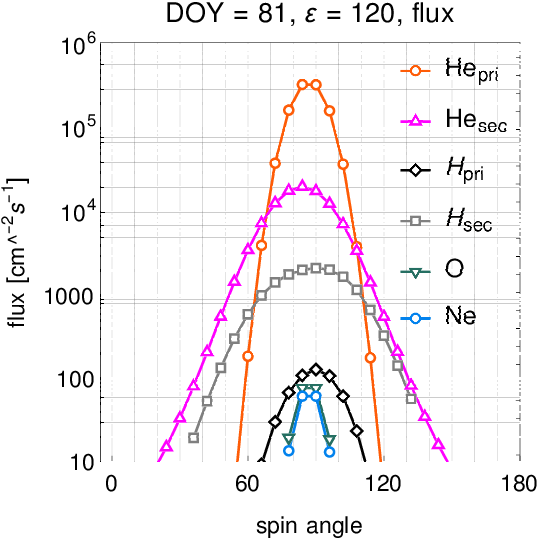}
\includegraphics[width=0.24\textwidth]{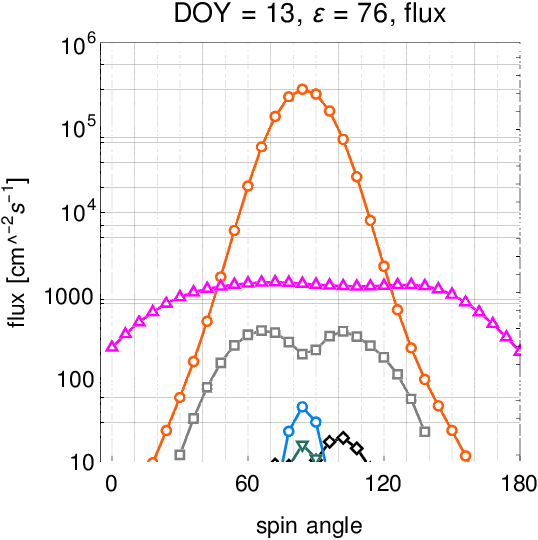}
\includegraphics[width=0.24\textwidth]{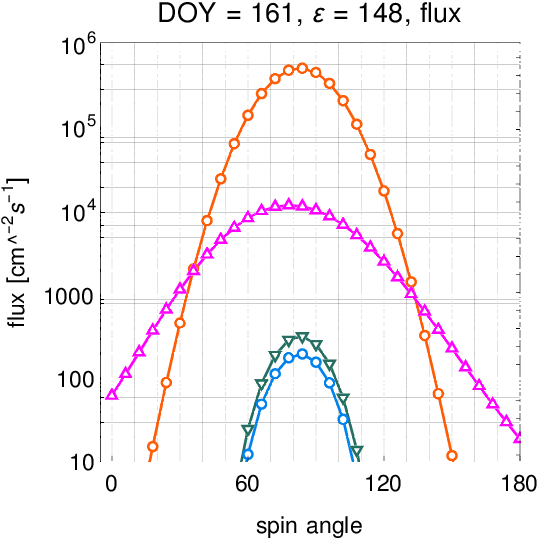}
\includegraphics[width=0.24\textwidth]{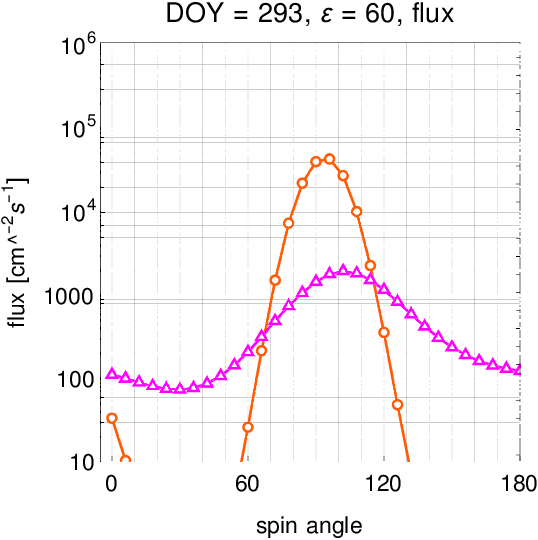}
\caption{
Illustration of the separation of the He signal from that of H. 
The first panel shows a (DOY, $\varepsilon$) combination for which a mixture of signals from both He and H populations is expected. 
In the second panel, the contribution from ISN H is several orders of magnitude smaller than that from He, including the far wings of the peak, where pure secondary He is visible.
In the third panel, there is no visible contribution from ISN H because the impact energy is below the detection threshold, and due to a small size of the scanning circle the count  statistics of He, Ne, and O is expected to be very good. 
The fourth panel presents an opportunity to see a mixture of indirect beams of the primary and secondary He without H and any other species. 
}
\label{fig:HebezH}
\end{figure}

\noindent
During solar maximum conditions, when the flux of ISN H is strongly reduced relative to that of ISN He, it is worthwhile to utilize this opportunity to maximize the observation time when pure He can be observed. The separation of the contributions to the measured signal from ISN He and ISN H is illustrated in Figure~\ref{fig:HebezH}. 

The first panel shows observations performed for (DOY = 81, $\varepsilon = 120\degr$). 
This is a typical situation, not favorable for unambiguous species separation because of a strong signal from H.
By contrast, for (DOY = 13, $\varepsilon = 76\degr$), shown in the second panel, the contribution from the secondary H is lower by an order of magnitude than in the former case, and the primary H is almost absent. 

Another parameter combination favorable for separation of H and He is (DOY = 161, $\varepsilon = 148\degr$). 
In this geometry, there is no H visible and, in addition to He, there is also a lot of O and Ne. 
However, because of the large difference in the energies, a contribution to the H$^-$ events registered by the instrument from Ne and O is negligible. 

Still another opportunity to observe He without H is shown in the fourth panel, where the signal is composed of the indirect beams of primary and secondary He. 
Here, no other species are expected in the signal, and hence this can be used for calibration of the instrument for He. Note, however, that radiation pressure might be affecting the signal, especially for the secondary population \citep{IKL:23a}. 

\subsection{Primary population of He without secondary He}
\label{sec:onlyHepr}
\noindent

The primary population of ISN He has a narrow beam relative to that of the secondary population.
This holds both for the spin angle (as shown in Figure~\ref{fig:HebezH}) and for the (DOY, $\varepsilon$) space, as shown in the upper two panels in Figure 7 in \citet{sokol_etal:19c}. 
Consequently, it is impossible to observe the primary He without any contribution from the secondary population. 
However, we identify some regions in the (DOY, $\varepsilon$) space where the secondary population is minimized, with the signal from the primary population exceeding that of the secondary populations by more than two orders of magnitude.
These regions correspond to the viewing geometry of IBEX. 
For IMAP, it is possible to achieve this effect for several lines in the (DOY, $\varepsilon$) space.

To successfully minimize the secondary population in the signal, it is important to carefully select the range of spin angles. It is recommended to follow the choice by \citet{bzowski_etal:15a}, \citet{swaczyna_etal:18a}, and \citet{swaczyna_etal:22b}, who selected 6 of the 6\degr{} spin angle bins, from 252\degr{} to 282\degr, that straddle the maximum of the signal. 
In the case of IMAP, when the elongation is different than $\varepsilon = 90\degr$, the signal is collected from a small circle in the sky. 
Consequently, for an elongation equal to, e.g., 148\degr, the number of the 6\degr{} spin angle bins is increased to 14, as shown (DOY = 161, $\varepsilon = 148\degr$) in the third panel of Figure \ref{fig:HebezH}. As an additional benefit, the expected counting statistics is better.

Another opportunity to observe the primary ISN He with little contribution from the secondary population focuses on the indirect beam. For example, for (DOY = 313, $\varepsilon = 60\degr$); see Figure \ref{fig:HePrSc}), the difference between the primary and secondary population fluxes is also larger than two orders of magnitude. 

\subsection{Secondary He without primary population}
\label{sec:onlyHeSc}
\noindent
The secondary population of ISN He is created near the heliopause and brings important information on the processes in the outer heliosheath. 
It is much weaker than the primary population, and in some regions of the (DOY, $\varepsilon$) space the separation is challenging.
Identification of regions in this space where the contribution from the primary ISN He is minimized would be advantageous. 

\begin{figure}
\centering
\includegraphics[width=0.24\textwidth]{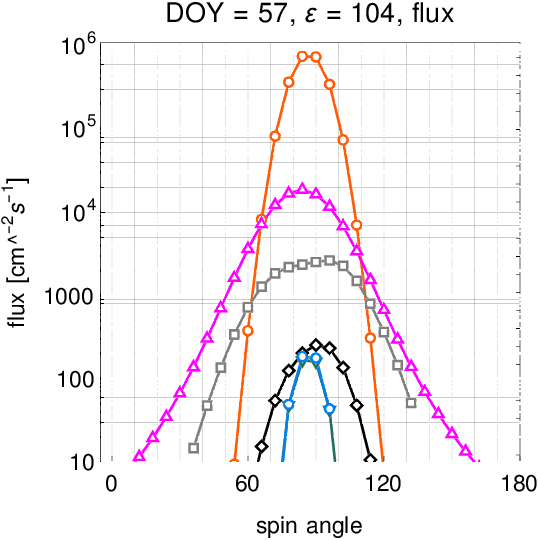}
\includegraphics[width=0.24\textwidth]{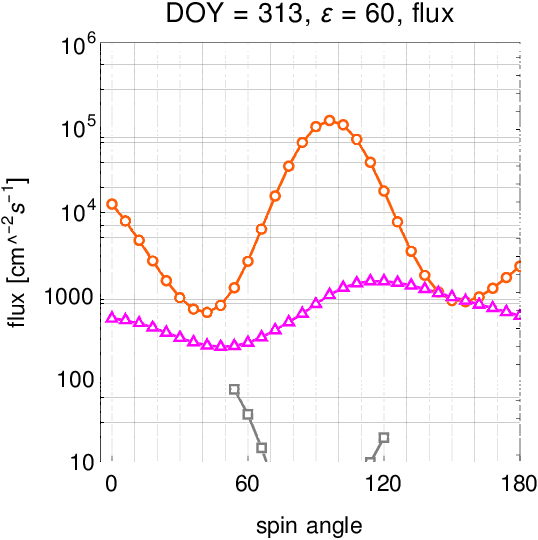}
\includegraphics[width=0.24\textwidth]{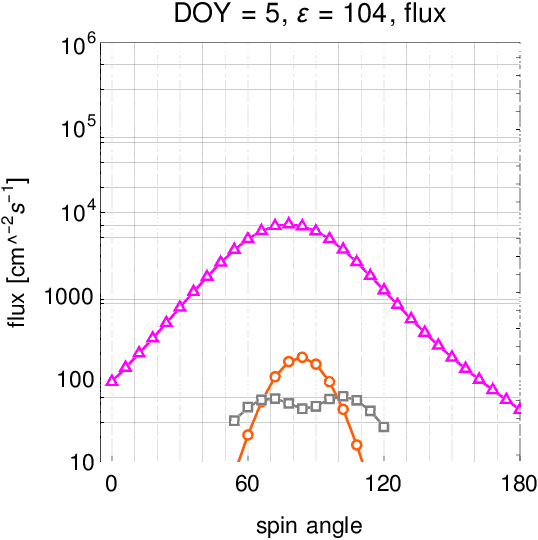}
\includegraphics[width=0.24\textwidth]{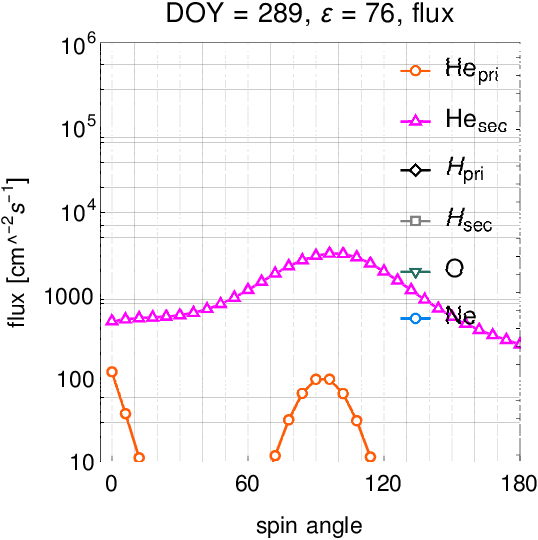}
\caption{
Sample illustration of (DOY, $\varepsilon$) combinations facilitating observations of the primary population of ISN He with a negligible contribution from the secondary population (first and second panel) and of the secondary population with a negligible contribution from the primary (third and fourth panel). 
Panel 2 shows a dominating primary indirect beam, with small contribution from the secondary indirect population. 
Panel 3 represents a viewing geometry where the direct beam of the secondary population is observed, and the primary population has not risen enough yet. 
The fourth panel represents a viewing geometry for the indirect beam of the secondary population.}
\label{fig:HePrSc}
\end{figure}

For IBEX ($\varepsilon = 90\degr$), the secondary population could be observed with little primary contribution during December and beginning of January. 
Depending on good observation times of the distribution during a given year, approximately 60 observation days were available, starting with the appearance of the secondary population in the field of view and ending when it was swamped by the primary population \citep{kubiak_etal:16a}.
Owing to the capability to change the elongation on IMAP, the time during the year when such observations are possible will be considerably longer. 
An example is presented in the panel for DOY 5 in Figure \ref{fig:HePrSc}. 
The primary population peak is visible already for $\varepsilon = 90\degr$, but it is still missing for $\varepsilon = 104\degr$. 
The magnitude of the secondary population flux is larger than that for the primary by two orders of magnitude ($\sim 10^4$ atoms \cmsq \persec). 
A similar situation is shown for the indirect beam, in the last panel in Figure \ref{fig:HePrSc}. 
For the smallest possible elongations, the signal for the secondary populations comes up first and increases much more rapidly than the primary signal. 
Here, we have favorable conditions for observing a pure secondary population. 
However, in this viewing geometry, interpretation of the signal requires taking the radiation pressure into account \citep{IKL:23a} which depends on the phase of the solar activity cycle.

It is also possible to take advantage of the fact that the secondary populations have much broader peaks as a function of spin angle than the primary population. 
Close to the passage through the maximum of its flux, e.g., at DOY 81 in Figure~\ref{fig:HebezH}, the secondary population features broad wings where there is no contribution from the primary population. 
In this case, it may be up to 6 bins on each side of the peak. 
  
There are more such regions in the (DOY, $\varepsilon$) space. They can be easily found by comparing the two upper panels of Figure 7 in \citet{sokol_etal:19c} (magenta dots). For example, the secondary population remains well visible for $\varepsilon = 136\degr$ throughout almost the entire year. Starting from DOY 180, it is the only population visible for this elongation. Thus, IMAP offers an excellent opportunity for detailed studies of the secondary population of ISN He and the related processes in the outer heliosheath. 

\section{Neon and Oxygen}
\label{sec:NeOx}
\noindent
One of the suggested ISN campaigns is that for the characterization of ISN O and Ne. The implementation of the campaign suggested in this paper offers extensive opportunities to characterize these species. We have a passage through the maximum of the flux, measurements of the anti-ram and indirect beams, and opportunities to minimize the contribution to the signal from O and thus observe pure Ne. For the first time, we discuss potential observations of the secondary O. 
The proposed elongation settings are also favorable for observations of ISN He, for which the ballistics is the same.

The observability limits for O and Ne that we used are different to those in \citet{sokol_etal:19c}: the minimum flux is 10 atoms \cmsq \persec, and the minimum number of spin angle bins above the threshold is 2. 
This is because the expected beams are very narrow. 
We stress that the magnitudes of the signal due to Ne and O we discuss here are comparable with the model values obtained for the geometry of the IBEX observations analyzed by \citet{schwadron_etal:16a}, which makes us believe that these species will also be observable for IMAP for the viewing geometries we discuss. 

\subsection{Flow parameters of the primary populations of O and Ne}
\label{sec:mainNeOx}
\noindent
The identification of O and Ne in the data is similar to that of separation of He and H. Oxygen atoms impacting the conversion surface capture an electron and are registered by the instrument as O$^-$ ions. 
Only a small fraction of the O atoms sputters O$^-$ and C$^-$ ions from the conversion surface. 
By contrast, Ne atoms do not produce Ne$^-$ ions and can only be registered due to O$^-$ and C$^-$ ions sputtered from the terrestrial water covering the conversion surface and the conversion surface itself. 
Thus, separation of the signals from Ne and O is challenging and can only be done statistically, on the ground. 

Assuming that the unperturbed velocities and temperatures of ISN O and Ne are identical, one expects the energies at the detector differing by the ratio of atomic masses of these species, i.e., by $\sim 1.25$. 
However, O produces a secondary component in the outer heliosheath \citep{park_etal:16a, baliukin_etal:17a}, and thus the flow parameters of its primary component may be modified. Conversely, Ne is not expected to produce a significant secondary population, and its velocity vector just inside the heliopause is expected to be very close to that in the unperturbed interstellar gas. 
It is thus interesting to independently determine the flow parameters of these species using the portions of data where one of the two species dominates.

\begin{figure}
\centering
\includegraphics[width=0.33\textwidth]{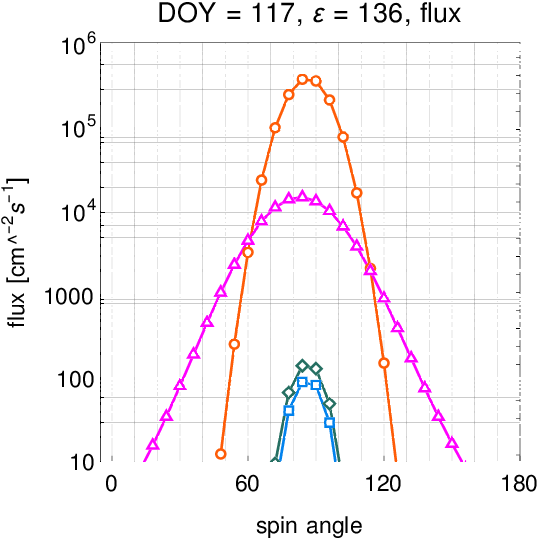}
\includegraphics[width=0.33\textwidth]{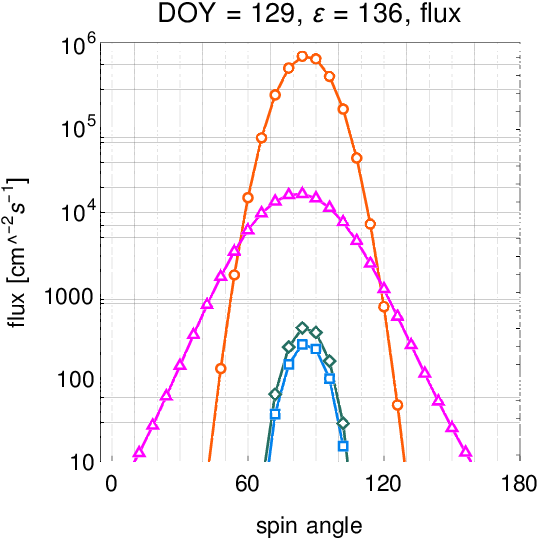}
\includegraphics[width=0.33\textwidth]{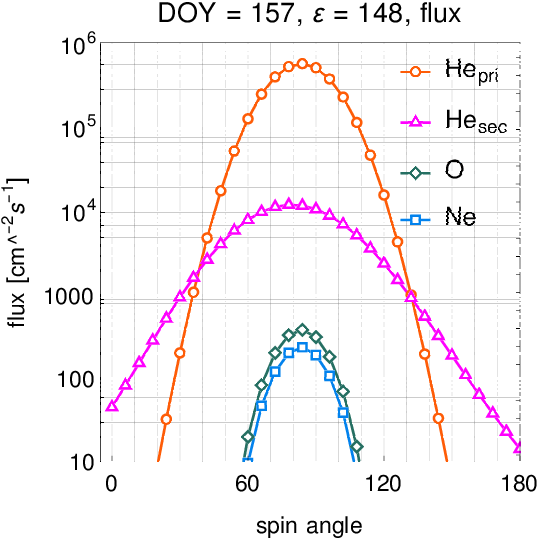}
\caption{
Simulated flux of the primary populations of ISN Ne and O and of the primary and secondary ISN He populations for a line of sight passage across the ISN beams with the elongation fixed at $\varepsilon = 136\degr$ (left and middle panels) and for $\varepsilon = 148\degr$ (right panel).
The left panel represents a viewing geometry where the beam is observed before reaching the maximum magnitude. 
The other two panels show viewing the beam for the elongation angles resulting in the maximum possible flux for the given DOYs.}
\label{fig:mainBeamNeOxPik}
\end{figure}

The strongest signal for Ne and O is expected during the first half of each calendar year.
Both of these species feature narrow beams, and consequently, for a given elongation, these beams are within the collimator field of view only for several days before they fade out.
Extending the observation time requires increasing the elongation angle. 
Within our proposed scenario of elongation angle settings, we can cross the beams of ISN Ne and O six times during a year (Figure~\ref{fig:obsPlan}), for different viewing geometries.
This offers IBEX-like measurements and also smaller and larger elongations. 
The setting $\varepsilon = 60\degr$ enables viewing the indirect beam, and $\varepsilon = 160\degr$ boosts the statistics because the scanning circle is small, and the beam is within the field of view during a large portion of the available observation time. 

For a fixed elongation, starting from the beginning of calendar year, the signal increases (see the progression of the peak heights in the first two panels of Figure~\ref{fig:mainBeamNeOxPik}). 
The daily maxima of the flux are presented in Figure~\ref{fig:mainBeamNeOxPik} for two elongations:  $\varepsilon=136\degr$ (middle panel), and $\varepsilon=148\degr$ (right panel). 
The width of the peak in spin angle increases with the increase in elongation because the larger the elongation, the smaller the angular radius of the scanning circle. 
This effect is clearly visible in the right panel for DOY 157. 
With an increasing elongation, the viewing time per spin angle bin increases, boosting the gathered number of counts per bin and improving the statistics. 
However, the correlation between spin angle bins is also increased because the angular width of the field of view becomes large in comparison with the angular arc of the spin angle bin.

\begin{figure}
\centering
\includegraphics[width=0.33\textwidth]{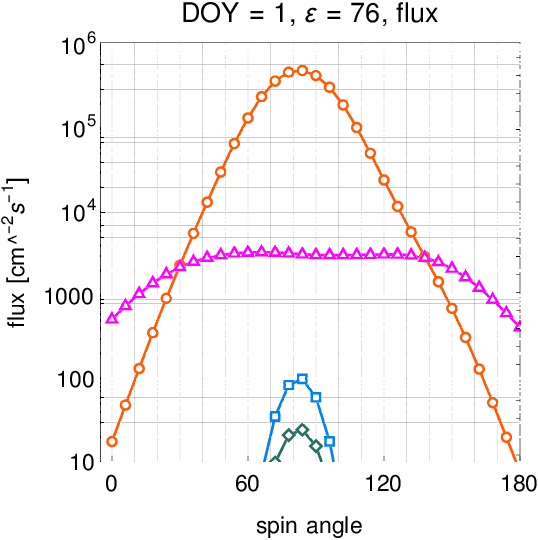}
\includegraphics[width=0.33\textwidth]{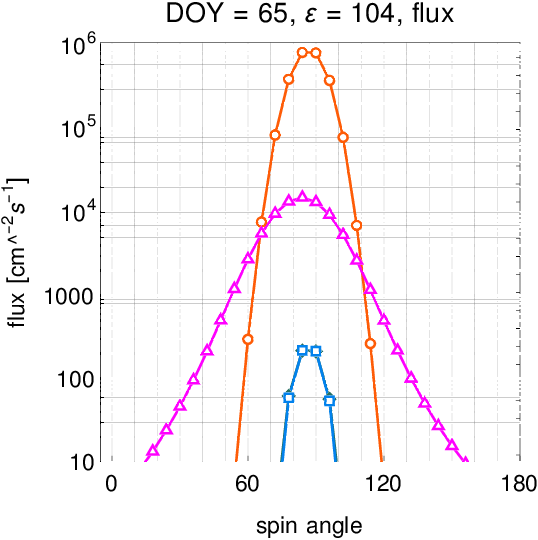}
\includegraphics[width=0.33\textwidth]{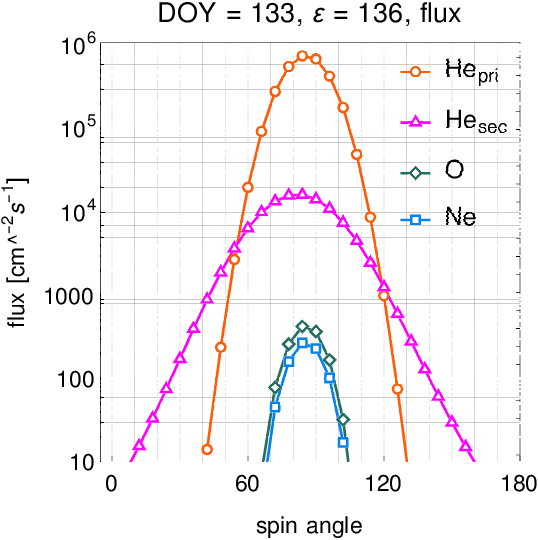}
\caption{
Opportunities for separating Ne and O. 
The panels show the expected fluxes for the primary O and Ne, as well as those for the primary and secondary He.
From left to right, they show the following conditions: Ne$>$O, Ne$=$O, Ne$<$O (the O flux larger than that of Ne by $\sim 70$\%.)
}
\label{fig:differNeOx}
\end{figure}

Since the viewing geometry favorable for O and Ne is also favorable for primary He, with appropriate energy setting of the instrument during the measurements, it will be possible to collect data usable for analysis of all three heavy species: He, Ne, and O. 
Examples of (DOY, $\varepsilon$) combinations for which large differences between the observed fluxes of Ne and O are expected are presented in Figure~\ref{fig:differNeOx} for the direct beam, and in Figure~\ref{fig:indirNeEner}, where only Ne is present in the indirect beam. 

The first panel of Figure \ref{fig:differNeOx} presents an opportunity to view Ne with little contribution from O. This observation can be used to verify the response of the instrument to Ne.
In the second panel, the magnitudes of the flux of Ne and O are almost identical, and the insight obtained from analysis of the first panel can help to differentiate between the two species. In the third panel, we have viewing conditions for which the contribution of O to  the total flux of Ne and O will be the greatest (O/Ne $\simeq 1.7$). 
These observations will be helpful in separation of the contributions from these two species to the signal -- see Figure 2 in \citet{bochsler_etal:12a}.

\subsection{Anti-ram beam of Ne: observations of Ne at different energies}
\label{sec:NeAntiram}
\noindent
By definition, in the ram geometry, the atoms come from the hemisphere containing the direction of the spacecraft velocity vector relative to the Sun.
This geometry is very favorable for ISN gas measurements because the velocities of the spacecraft and the atoms add up. Thus, this situation results in larger impact energies, and consequently higher detection efficiencies.
The anti-ram geometry corresponds to the opposite situation.

\begin{figure}
\centering
\includegraphics[width=0.33\textwidth]{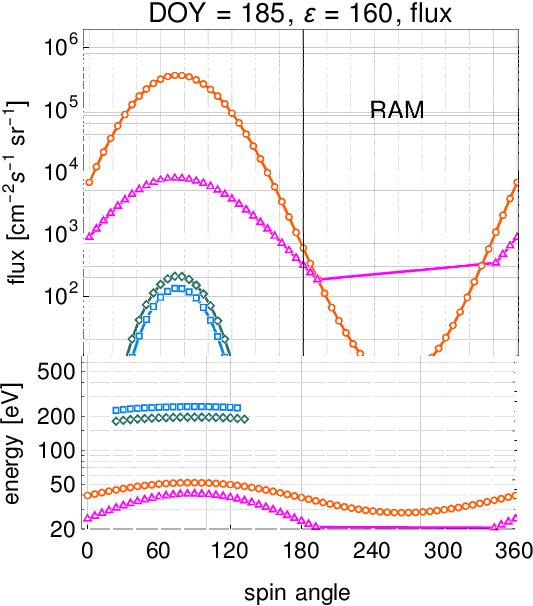}
\includegraphics[width=0.33\textwidth]{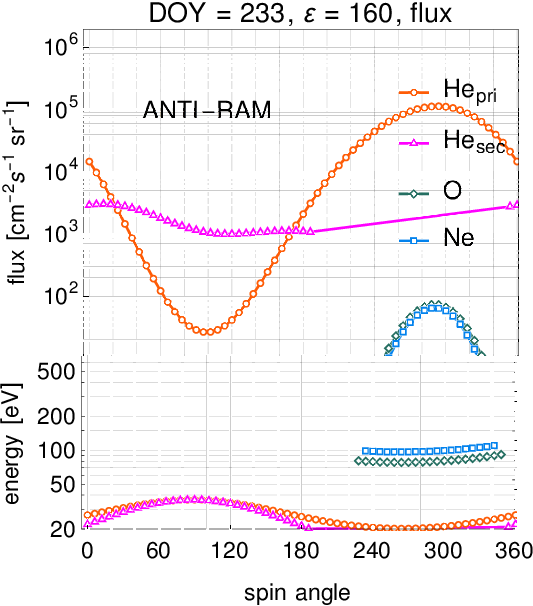}
\caption{
Expected fluxes and impact energies of ISN He, Ne, and O for two viewing geometries when the spacecraft is close to the upwind direction.
The upper portions of the two panels present the fluxes, and the lower portions the corresponding impact energies.
The left panel presents detection in the ram viewing geometry, with the energies of all species above the detection threshold of 20 eV.
The right panel corresponds to a situation when the spacecraft has already passed the upwind direction and the atoms are chasing the spacecraft, i.e., to the anti-ram geometry. 
Only the O and Ne components exceed the energy detection threshold and feature a flux sufficiently high to expect their detection. 
The flux of ISN He is high but its energy for the primary population barely exceeds the detection threshold, and that of the secondary He is below it.
}
\label{fig:antiRamNeOx}
\end{figure}

Observations of ISN He for the anti-ram viewing geometry have been challenging because of the low impact energy of the atoms at the detector. 
However, for the heavy Ne and O species, the energy is likely to be sufficient for successful detection even in the anti-ram geometry, that are only accessible at elevation angles substantially different from $90\degr$. 
In Figure~\ref{fig:antiRamNeOx}, we compare the simulated fluxes for the ram and anti-ram geometries. The former is shown for DOY 185, and the latter for DOY 233, both for $\varepsilon = 160\degr$, which is the limiting elongation for IMAP-Lo. 
In both of these cases, the beam is direct. The spacecraft is in the upwind region of its orbit around the Sun.
For the first of these two combinations, the energies are much larger than 200 eV for both Ne and O, and for the second combination, the energies are close to 70 eV. 
Comparing the results from these two combinations is favorable for verification of the instrument calibration and of the existence of a possible energy threshold for detection of the heavy species.

Since the observations are performed for upwind vantage locations, the absolute densities of these species are the least modified by ionization among all possible viewing geometries accessible to IMAP-Lo. 
For He, the impact energies are too low to expect atom detection in the anti-ram viewing geometry, shown in the second panel. 
The large elongation of 160\degr{} makes the scanning circle very small, and the statistics will be improved because the beam is within the detector field of view for a large portion of the observation time.

\subsection{Indirect and mixed beams of Ne}
\label{sec:Neindir}
\begin{figure}
\centering
\includegraphics[width=0.33\textwidth]{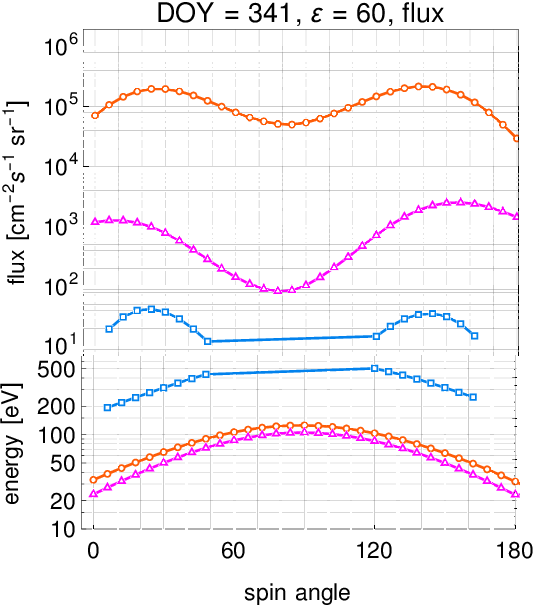}
\includegraphics[width=0.33\textwidth]{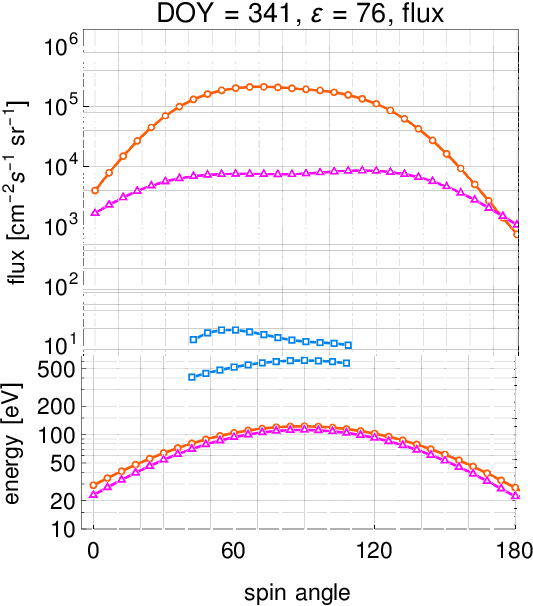}
\includegraphics[width=0.33\textwidth]{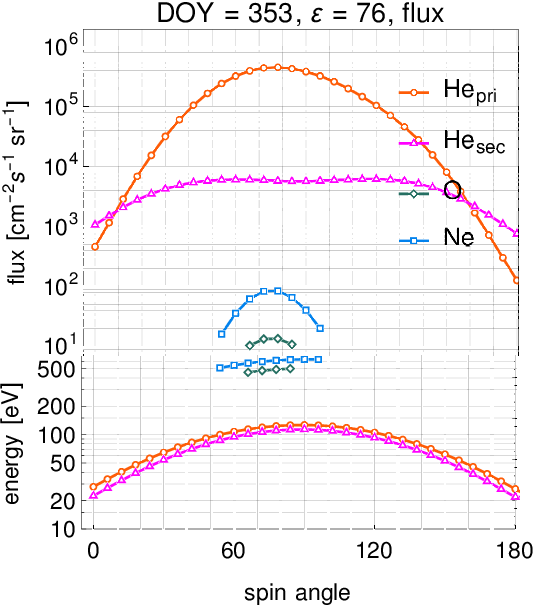}
\caption{
Opportunities for studies of ISN Ne using small elongation angles. 
The upper portions of the panels present the fluxes, the lower portions the respective impact energies (color-coded).
The left panel illustrates an opportunity to view ISN Ne with no contribution from ISN O.
Both the direct and indirect beams are in the instrument visibility strip, and the O atoms are missing because they suffered much larger ionization losses than the Ne atoms.
Also the primary and secondary populations of ISN He are visible, but in their case, a clear separation between the direct and indirect beams is not easy because of the larger thermal spread of He.
The second panel presents the same DOY but a larger elongation angle (76\degr{} vs 60\degr). This scenario corresponds to a viewing geometry where both the He and Ne beams have merged but the expected Ne signal is close to the adopted flux detection threshold. 
The third panel shows a (DOY, $\varepsilon$) combination for which only the direct beam is in the visibility strip and both ISN O and Ne are visible. 
A combination of the viewing geometries represented in the first and last panels can be used to verify the calibration of the instrument for O and Ne atoms.}
\label{fig:indirNeEner}
\end{figure}

\noindent
Successful measurement of the indirect beam of ISN Ne is interesting, among others, because it facilitates an independent measurement of the ionization rate of ISN Ne, as discussed by \citet{bzowski_etal:23a}. 
Due to the high impact energy of ISN Ne atoms, there is a short time interval during the year near DOY 341 when it is possible to observe both the direct and indirect beams of ISN Ne simultaneously. 
The indirect beam is shown in the left panel of Figure~\ref{fig:indirNeEner}, where it is visible close to 150\degr{} in spin angle. 
The direct beam appears close to 30\degr{} in spin angle.
Both of these beams exceed the flux and energy detection thresholds. 
This combination of (DOY, $\varepsilon$) offers a rare opportunity to observe the direct and indirect beams simultaneously.
Furthermore, the energies of the two beams are almost identical, as shown in the lower part of the panel. 

\citet{bzowski_etal:23a} did not include this combination of (DOY, $\varepsilon$) recommended for studies of the ionization rate of ISN He, even though the direct and indirect beams of ISN He are also visible. 
This was because in the case of ISN He, which has a much larger thermal spread, a clear separation of the direct and indirect beams is not easy.
However, our simulations show that if the two Ne beams are detected, there will be no problems with separating them into the direct and indirect beams, so potentially, this combination of (DOY, $\varepsilon$) may be useful for independent determination of the ionization rate of ISN Ne.
Additionally, no ISN O contribution is expected in this viewing geometry, so it is possible to examine the sensitivity of the instrument to pure Ne and facilitate the interpretation of the measurements where both Ne and O are present. 

Setting the instrument to $\varepsilon = 76\degr$ offers a different view of the same beams. 
The expected Ne signal is much weaker, and the direct and indirect beams are merged, as shown in the center panel of Figure \ref{fig:indirNeEner}. 
In our scenario, we suggest exercising both of these (DOY, $\varepsilon$) combinations. 

The third panel of Figure \ref{fig:indirNeEner} presents the direct beam of ISN Ne, with a $\sim 10$\% contribution of primary O to the flux. 
Together with the data collected for the (DOY, $\varepsilon$) combination illustrated in the first panel of the figure, this offers an opportunity to better verify the instrument calibration for O and Ne.
This viewing geometry is suggested for both years of observations.

\begin{figure}
\centering
\includegraphics[width=0.4\textwidth]{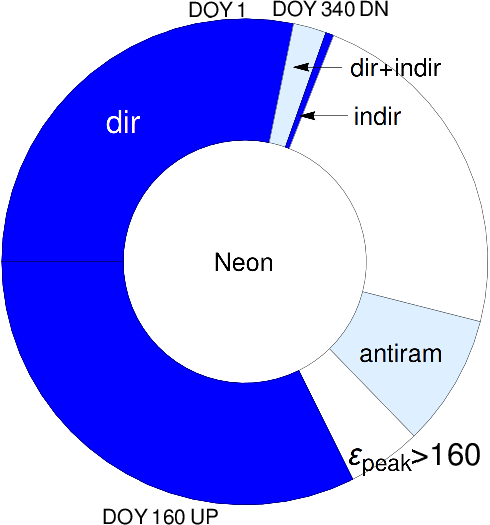}
\caption{
A pie chart of the observation times for ISN Ne over the calendar year. See text for details.}
\label{fig:NeTimes}
\end{figure}

A summary of this section is presented in Figure~\ref{fig:NeTimes}.
ISN Ne is  observable for more than half of the year. 
Different times during the year require different settings of the elongation angle.
The dark-blue region corresponds to the direct beam of the primary ISN Ne observed in the ram geometry.
The region requiring elongation $\varepsilon > 160\degr$ is not accessible.
Subsequently, another opportunity for observing ISN Ne appears, a month long, when the direct beam can be observed in the anti-ram geometry, with inevitably lower impact energies.
When this interval is over, there is a time when ISN Ne cannot be observed by IMAP-Lo.
Around DOY 337, the indirect beam of ISN Ne becomes visible.
This interval is short, just several days (see the thin dark-blue strip in Figure \ref{fig:NeTimes}).
Shortly thereafter, a week long opportunity to view the cone of ISN Ne appears, where the direct and indirect beams become indistinguishable.
This is marked in pale blue.
After this one, IMAP can return to observation of the direct beam in the ram geometry.

\subsection{Secondary oxygen population}
\label{sec:OxSec}
\noindent
The secondary population of ISN O has been detected in the IBEX data. 
The flow parameters of this population are known with large uncertainties. \citet{park_etal:16a} and \citet{park_etal:19a} discovered the secondary O population in the IBEX-Lo data and suggested its inflow direction as longitude $247\degr \pm 1.5\degr$, latitude $12\degr \pm 1.6\degr$, with the inflow speed $11 \pm 2.2$ \kms{} and a temperature of $(10 \pm 1.5) \times 10^3$ K. 
These values result in a deflection of 4\degr{} from the inflow direction of the secondary He population reported by \citet{kubiak_etal:16a}. 
This flow direction is inclined at an angle of 2\degr{} to the B-V plane determined by the interstellar magnetic field vector obtained by \citet{zirnstein_etal:16b} and the inflow direction of ISN He from \citet{bzowski_etal:15a}.
The density of the secondary O population was estimated as 50\% of the density of the primary ISN O.

\begin{figure}
\centering
\includegraphics[width=0.6\textwidth]{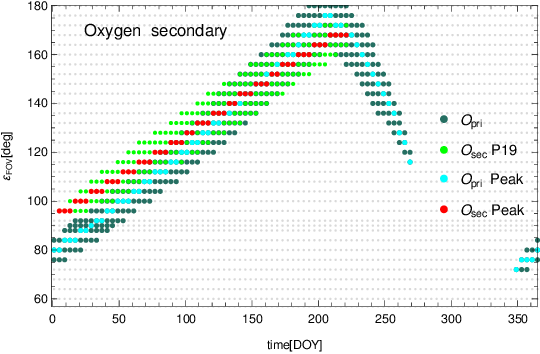}
\caption{
Opportunities for viewing the secondary population of ISN O. 
The figure is in a similar format to that used in Figure 7 by \citet{sokol_etal:19c}. 
Dark-green dots (\oxpri) represent the combinations of (DOY, $\varepsilon$) for the primary population of ISN O. 
The light-green dots represent the secondary population of ISN O with the inflow parameters from \citet{park_etal:19a} (\oxsec{} P19 in the figure) and the density equal to 50\% of that of the primary assumed. 
The cyan (\oxpri{} Peak) and red dots (\oxsec{} Peak) mark the positions of the maximum of the flux for the primary and secondary populations, respectively. 
To qualify for this map, the flux must be larger than 10 atoms \cmsq \persec{} for at least three spin angle bins. }
\label{fig:OxsecdoyElong}
\end{figure}
\begin{figure}
\centering
\includegraphics[width=0.24\textwidth]{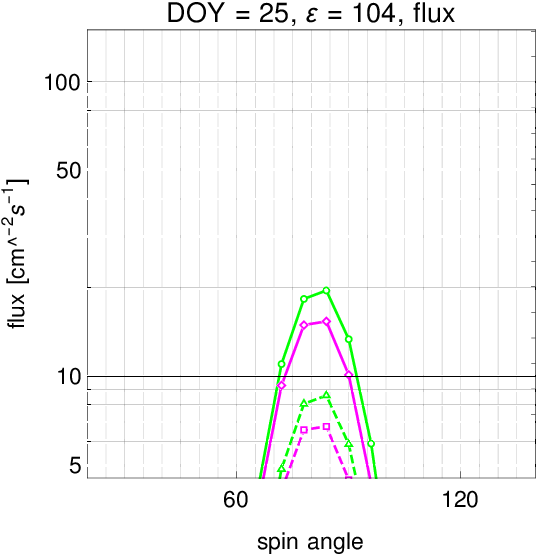}
\includegraphics[width=0.24\textwidth]{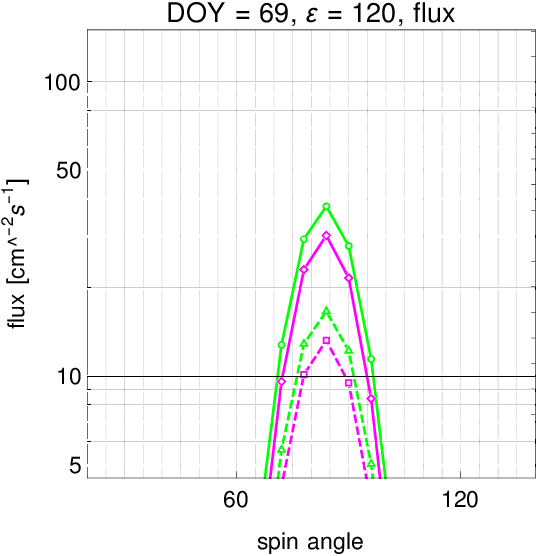}
\includegraphics[width=0.24\textwidth]{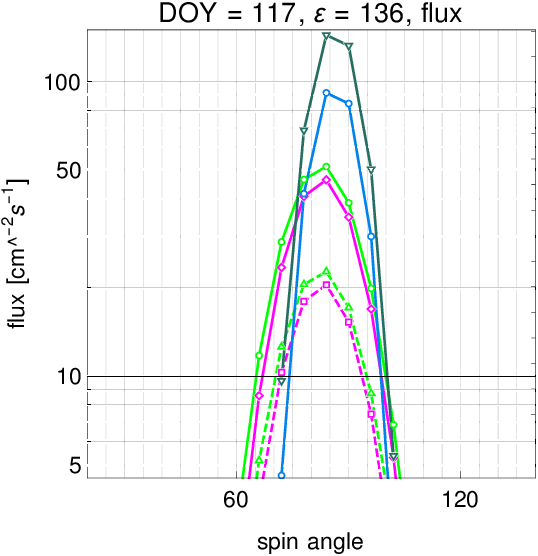}
\includegraphics[width=0.24\textwidth]{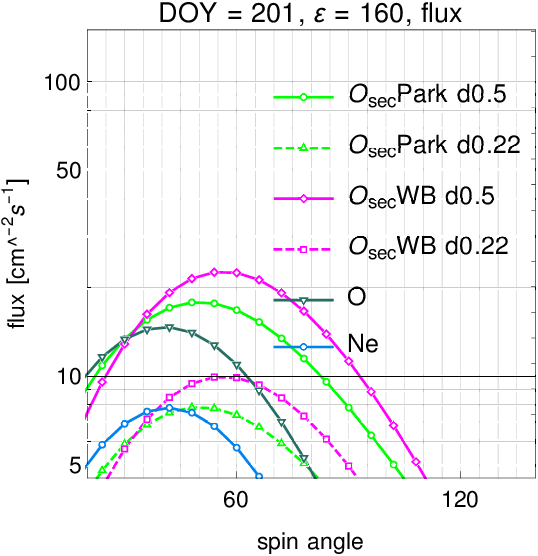}
\caption{
Simulated fluxes of the primary and secondary ISN O and of the primary Ne for (DOY, $\varepsilon$) combinations well suitable for observations of the secondary O. 
Four cases of hypothetical parameters of the secondary O populations are displayed. 
The simulations for the two hypothetical densities are presented with solid (for the density of 50\% of that of the primary) or broken lines (for 22\%). 
The green vs magenta colors differentiate between the flow parameters from \citet{park_etal:19a} and \citet{kubiak_etal:16a}.
The two panels at the left side show the times when the primary O and Ne are not yet visible. 
The third panel represents an example when the secondary population is well visible at the wings of the ISN O peak. 
The extreme right panel illustrates a situation where the primary populations have already disappeared from the field of view, but the secondary populations are still well visible.}
\label{fig:Oxsec}
\end{figure}

\citet{baliukin_etal:17a} simulated the filtration of the primary ISN O through the outer heliosheath and production of the secondary O. They concluded that the density of the secondary population is about 22\% of the density of the primary population.
However, the input parameters to the global heliosphere model used by these authors differed from those obtained from direct-sampling observations of ISN He and of the interstellar magnetic field vector obtained from the IBEX Ribbon analysis by \citet{zirnstein_etal:16b}. 

In this paper, we identify the region in the (DOY, $\varepsilon$) space best suitable for measuring the secondary O population within the adopted observation scheme. 
We assumed that the inflow parameters for the secondary O are identical to those for the Warm Breeze (i.e., the secondary He) and performed the simulations for two densities: either 50\% of that for the primary ISN O at the heliopause, as suggested by \citet{park_etal:19a}, or 22\%, as suggested by \citet{baliukin_etal:17a}. 
Alternatively, we simulated the secondary O for the flow parameters suggested by \citet{park_etal:19a}, also for the two alternative densities. 

The simulations were performed for the same (DOY, $\varepsilon$) grid as that used so far, and for the same ionization rate model \citep{sokol_etal:19a}. 
Because of the large uncertainties of the inflow parameters, this study must be regarded as a qualitative reconnaissance rather than a prediction of results of future observations.
Even though the flow parameters of the secondary populations of O and He were adopted as identical, the observed flow pattern of the two populations potentially may differ because of the different thermal spreads and very different ionization rates.

We start the analysis from looking at the entire (DOY, $\varepsilon$) space for 2015, as shown in Figure \ref{fig:OxsecdoyElong} in a similar format to Figure 7 in \citet{sokol_etal:19c}. 
We present the prospective (DOY, $\varepsilon$) combinations for observations of the secondary O for the Park parameters with the assumed density equal to 50\% of that of the primary ISN O at the heliopause. 
To qualify for the map, we request that the flux of a population exceeds 10 atoms\cmsq \persec{} for at least 3 spin angle bins. 

Inspection of the figure suggests that there are quite substantial regions in the (DOY, $\varepsilon$) space where the secondary O might be observable (light green dots). 
However, the key factor here is the density: if it is equal to 22\% of that of the primary, as suggested by \citet{baliukin_etal:17a}, then the detection opportunities are somewhat reduced. 
Then, the secondary O becomes visible only around DOY 30 and fades out 30 days earlier than in the alternative case with 50\% density, suggested by \citet{park_etal:19a}. 
Still, it offers more than 150 days of observations. 
It is worthwhile to look into the indicated regions and collect good counting statistics because detection or non-detection makes a simple test for the density of the secondary O.

In this context, it is important to perform the observations suggested earlier in the paper that verify the signatures of Ne vs O in the detector, i.e., to perform observations when only Ne is expected to be visible. 
This is to provide means of verification if a signal observed for the times, viewing geometries, and spin angle ranges where \oxsec{} is expected is indeed due to O atoms. This distinction is important because we expect to detect secondary O but not secondary Ne. 
Detection of Ne atoms in the region where only secondary O is expected would be an important discovery. 

Figure \ref{fig:Oxsec} presents details of the flux of the secondary O as a function of spin angle. 
Different colors mark different flow parameters, and the solid vs broken lines correspond to the two alternative density values.

Potential opportunities to view the secondary O population appears during the days preceding those favorable for observations of the primary ISN O.
For a given elongation setting, the secondary population appears first, as in the case of the secondary He, and the primary population comes up later. 
This situation is illustrated in Figure \ref{fig:Oxsec}, for two elongations: $\varepsilon = 104\degr$ and $\varepsilon = 120\degr$.
Both of the presented (DOY, $\varepsilon$) cases are for a time before the maximum of the ISN O peak appears.

Even when the primary ISN O is present in the signal, it is possible to observe the secondary O at the wings in the spin-angle distribution of the observed flux, as illustrated in the third panel in Figure \ref{fig:Oxsec}, similar to the case of the secondary He.

For differentiation between the various parameter sets of the secondary ISN O, the favorable conditions are obtained for times near DOY 201, as shown in the right panel of Figure \ref{fig:Oxsec}.
The recommended elongation is $\varepsilon = 160\degr$, which offers a long exposure time due to the small radius of the scanning circle -- and consequently a boost for the count statistics. 
This combination is favorable because the four different hypotheses on the flow parameters and densities for the secondary O offer different flux vs spin angle relations.
Additionally, the contribution from Ne is negligible, and that from primary O is relatively small. 
For the first and second panel of Figure~\ref{fig:Oxsec} the same is true, but the expected count rates are lower and thus the statistics poorer. 

\section{Summary and conclusions}
\label{sec:conclu}
\noindent
Observations of various species and populations of ISN atoms are a potential plentiful source of information on interstellar matter surrounding the Sun and processes operating in the outer heliosheath.
This information can be obtained owing to the capability of the IMAP-Lo instrument to vary the radius of its Sun centered scanning circle. 
This paper is a continuation of a line of papers discussing various science aspects of the future observations of the ISN gas by IMAP-Lo: \citet{sokol_etal:19c, schwadron_etal:22a, bzowski_etal:22a, bzowski_etal:23a}, which use the same simulation basis.
 
Based on extensive simulations of ISN beams in various viewing geometries performed using the WTPM simulation code, we devised a scenario of setting the radius of the instrument scanning circle, referred to as the elongation angle, that allows to efficiently address multiple science opportunities given by the ability of the instrument to observe ISN H, D, He, Ne, and O, with their primary and secondary populations. 
In this paper, we present this scenario focusing on the heavy species (He, O, and Ne). A forthcoming paper will discuss H and D in greater detail.

The simulations used in the paper are supposed to offer illustrative, realistic examples, but because the ionization rate model used is not up to date, and it is impossible to predict the ionization rates for the dates of actually performed observations, they are not predictions of the observed signal.

From the operational point of view, we suggest organizing observations in four-day cycles. 
During days A and C of each cycle, the focus is put on studies of the ISN species, and during days B and D, on observations of the ENAs.
Even though the proposed scenario differentiates between the ISN and ENA days, the observations collected during ENA days can be successfully used in ISN analysis, and vice versa.

The complement of suggested observations can be accomplished during two consecutive observation years.
The observations can start at an arbitrary DOY, but the elongation setting scenario is strictly connected with the DOY because of geometry reasons.

During the first 365 days of observations, it is suggested to observe the ISN atoms at elongations 75\degr{} and 105\degr{} alternately, and the ENAs at a fixed elongation of 90\degr.
This results in performing observations along three parallel lines in the (DOY, $\varepsilon$) space, as illustrated in the first panel of Figure \ref{fig:obsPlan}. 

During the second year of observations, the ENAs are observed at $\varepsilon$ angles alternating between 90\degr{} and 135\degr. 
The ISN atoms are observed along parallel lines in the (DOY, $\varepsilon$) space shown in the second panel of Figure \ref{fig:obsPlan}, with the boundaries defined in Table \ref{tab:Y2}.
The proposed lines of elongations allow for multiple crossings of the beams of all of the ISN species visible to IMAP-Lo.
It is understood that small differences in the elongation angle settings are possible without loss of the science opportunities, e.g., adopting $\varepsilon$ alternating between 75\degr{} and 105\degr{} instead of 76\degr{} and 104\degr{} during the first year, or 150\degr{} instead of 148\degr{} during the second year. 
We used the values presented in the paper for the sake of homogeneity with the common simulation basis, used also in the other papers within this topic of interest.

From the perspective of the suggested observations and their analysis, the data are split into several topical campaigns, discussed in Section \ref{sec:obsPlan} and illustrated with a color code in Figure \ref{fig:obsPlan}.
This sequence allows to obtain observations of individual populations (primary without secondary and secondary without primary), as well as to view ISN Ne without contribution from ISN O, to look for secondary O, and to view the indirect beams of ISN He (both primary and secondary).
Based on these observations, it will be possible to break the correlation of the flow parameters of the ISN gas, to determine these parameters separately for He, O, and Ne, to do this also for the secondary He, and to efficiently search for the secondary population of ISN O.
It will also be possible to determine the ionization rate of ISN He and possibly Ne without a calibration bias. 

The suggested scenario enables accomplishing all of the science goals for the IMAP-Lo instrument and offers a viable way of doing it within the first two years of the mission.
This scenario will be utilized as a tool to complement ongoing IMAP planning work. Possible improvements may be studied in the future to account for solar cycle variations if scenarios are considered for other phases of the solar cycle than solar maximum. 

\begin{acknowledgments}
\emph{Acknowledgments} We are obliged to Fatemeh Rahmanifard and Stephen Fuselier for stimulating discussions. The work by M.A.K. and M.B. was supported by the National Science Centre, Poland (grant 2019/35/B/ST9/01241). 
P.S. acknowledges support from the Polish National Agency for
Academic Exchange within Polish Returns Programme (BPN/PPO/2022/1/00017) and the
National Science Centre, Poland (grant 2023/02/1/ST9/00004). 
The work by E.M., N.S, and D.M. was funded by the IMAP mission as a part of NASA's Solar Terrestrial Probes (STP) Program (80GSFC19C0027).
\end{acknowledgments}

\appendix
\section{Fluxes and corresponding energies for all four-day cycles within the proposed scenario}
\label{sec:appendix}
\begin{figure}
\centering
\includegraphics[width=0.75\textwidth]{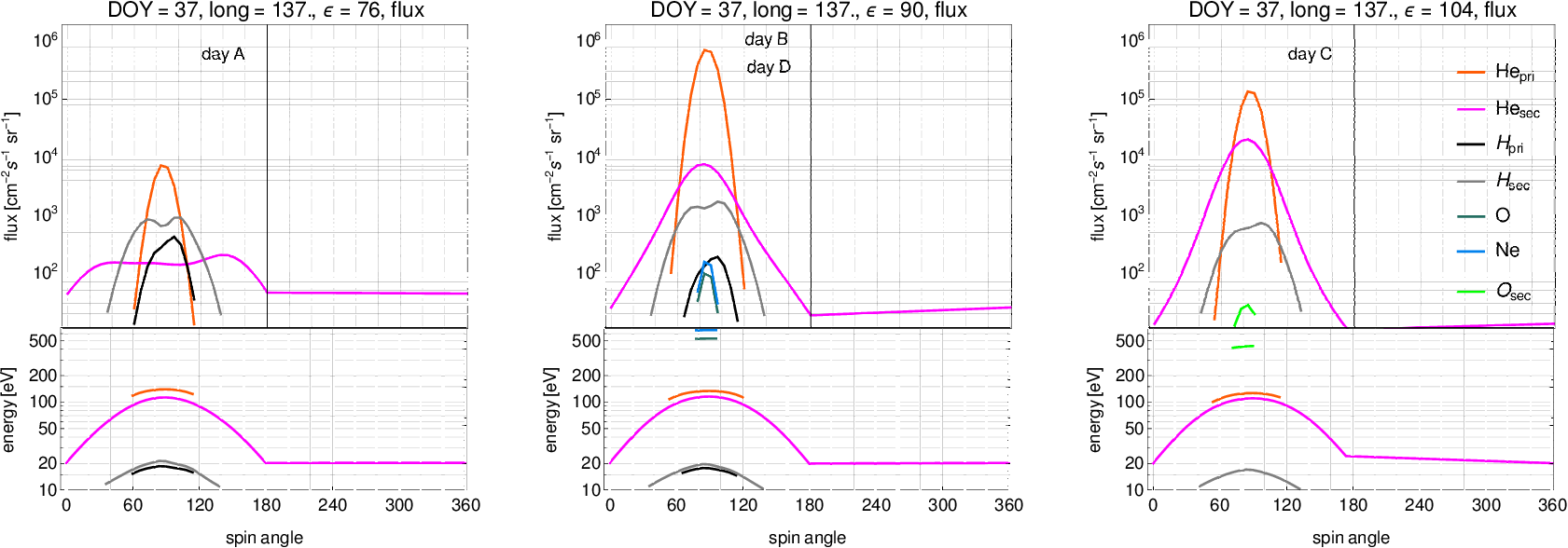}
\caption{
Example Figure for figure set. Fluxes and corresponding energies of all relevant species and their populations for individual four-day observation cycles for the first year of the proposed scenario.
The species and populations are marked up in the legend within the right panel. 
The DOY and elongation, as well as the corresponding ecliptic longitude of the observer, are shown in the panel headers. 
The respective day within a cycle (A, B, C, or D) are indicated within the panels. 
The horizontal axes are the IMAP spin angles, the vertical axis for the flux are given in absolute units -- without any geometric factors applied.
The vertical axis for the energies is in eV; these are absolute energies, with specific masses of individual species included.
This is an example of a series of electronic figures available on request }
\label{fig:opsPlanFluxes1Y}
\end{figure}

\begin{figure}
\centering
\includegraphics[width=\textwidth]{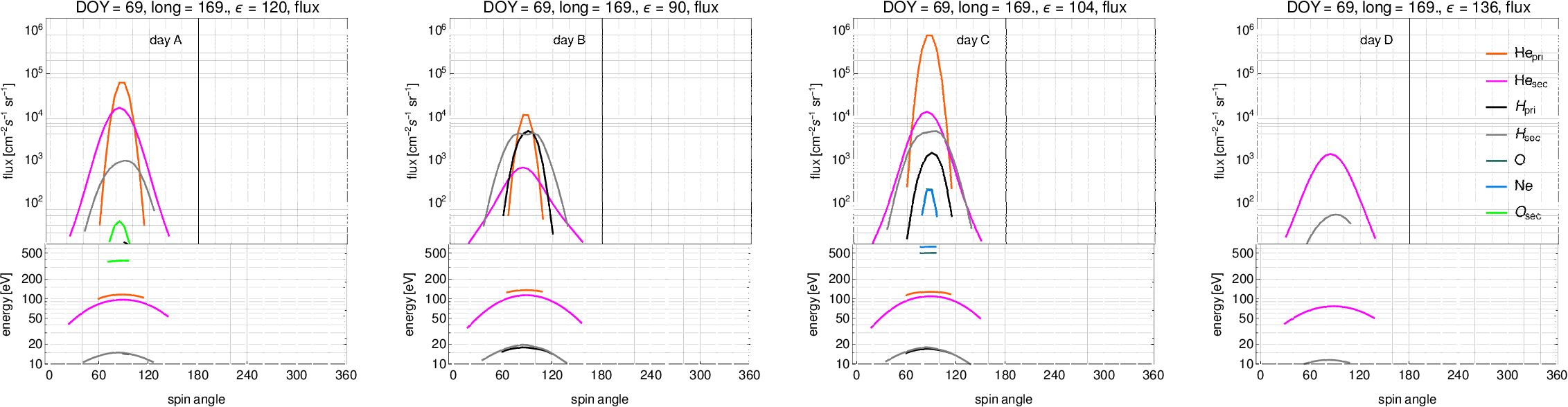}
\caption{\emph{opsPlanFluxes2Y} 
Same as Figure \ref{fig:opsPlanFluxes1Y}, but for the second year within the proposed scenario. 
Since during this year, we suggest to always follow four parallel lines in the (DOY, $\varepsilon$) space, we have four different settings within a cycle, represented in four panels.
This is an example of a series of electronic figures available on request. }
\label{fig:opsPlanFluxes2Y}
\end{figure}

\noindent
In this Appendix, we present simulations of the fluxes of all relevant ISN species and their populations for all 4-day cycles within our elongation setting scenario. 
Year 1 is displayed in the  example in Figure \ref{fig:opsPlanFluxes1Y}, and Year 2 in Figure \ref{fig:opsPlanFluxes2Y}.

In both of these figures, we present the fluxes and corresponding impact energies for individual 4-day cycles of observations. 
The upper portions of the panels represent the fluxes, and the lower ones the corresponding energies, averaged within individual spin angle bins. 

The elongation angle settings for the first year alternate between 76\degr{} and 104\degr{} for the ISN days (A and C) and 90\degr{} for both of the ENA days (B and D) within a cycle. 
Therefore, we represent the fluxes in three panels for each of the cycles, with the second panel corresponding to $\varepsilon = 90\degr$.

For the second year, the ENA days alternate between 90\degr{} and 136\degr{} and are represented in the second and fourth panels. 
The elongation for the ISN days varies according to the scenario presented in Figure \ref{fig:obsPlan} and Table \ref{tab:Y2}.
In both figures, the DOYs, corresponding ecliptic longitudes of the observer, and elongation angles are displayed in the panel headers, and the day within a cycle (A---D) is indicated inside the panels.

The fluxes are presented with the thresholds applied. 
To qualify for display, the flux within an individual 6\degr{} spin angle bin for a given population must exceed the flux threshold, and also the minimum energy threshold. 
The magnitudes of these thresholds are discussed in the main text.

Since we only have simulations for every fourth DOY, we represent the fluxes within a given cycle simulated for the first DOY of the given cycle. 
It is expected that changes in the fluxes due to the spacecraft motion around the Sun by $\sim 4\degr$ in ecliptic longitudes are small enough to warrant this approximation.

\bibliographystyle{aasjournal}
\bibliography{iplbib}
 
\end{document}